\documentclass[a4paper,11pt]{article}[margin=1cm]

\pdfoutput=1 
\usepackage{jcappub} 
\usepackage[T1]{fontenc} 
\usepackage{multirow}
\usepackage{tikz}
\usepackage{multicol}
\usepackage{graphicx}
\usepackage{txfonts}
\usepackage{amsmath}
\usepackage{tabularx}   
\usepackage{dcolumn}    
\usepackage{booktabs}   
\usepackage{siunitx}    
\newcolumntype{d}{D{.}{\times}{-1}} 
\usepackage{hyperref}
\usepackage{todonotes}
\usepackage{subcaption}
\usepackage{float}
\usepackage{orcidlink}
\usepackage[sort&compress]{natbib}
\usepackage[utf8]{inputenc}
\DeclareUnicodeCharacter{2009}{\thinspace}
\hypersetup{colorlinks,linkcolor={blue},citecolor={blue},urlcolor={blue}}

\title{
Discovering the Dispersion of Gravitational Waves using Multi-Band Observation including Deci-Hertz: A Unique Probe to  Cosmic Acceleration}

\author[a]{Adith Praveen,\orcidlink{0009-0002-3237-2960}}
\author[b]{Panchanjanya Dey,\orcidlink{0009-0006-0270-7394}}
\author[c]{Suvodip Mukherjee \orcidlink{0000-0002-3373-5236}}

\affiliation[a]{Department of Physics, Indian Institute of Technology Madras, Chennai-600036, India}
\affiliation[b]{Department of Physics, Indian Institute of Science, Bangalore-560012, India}
\affiliation[c]{Department of Astronomy and Astrophysics, Tata Institute of Fundamental Research,\\1 Homi Bhabha Road, Colaba, Mumbai-400005, India}

\emailAdd{ph21b001@smail.iitm.ac.in}
\emailAdd{panchajanyad@iisc.ac.in}
\emailAdd{suvodip@tifr.res.in}

\abstract{
The dispersion in the speed of gravitational waves is a novel way to test the general theory of relativity and understand whether the origin of cosmic acceleration is due to any alternative theory of gravity. Several alternative theories of gravity predict dispersion in the gravitational wave signal in a frequency-dependent deviation from the speed of light at lower frequencies than accessible from current ground-based detectors.   {We show how a multi-band observation of gravitational wave signal combining deci-Hertz gravitational wave signal from LGWA (Lunar Gravitational Wave Antenna) with ground-based detectors such as Cosmic Explorer or Einstein Telescope, and also including LISA (Laser Interferometer Space Antenna), we can probe the energy scale associated with effective theory of modified gravity scenarios by combining only $\mathcal{O}(10)$ high  {signal to noise ratio} (SNR) with a precision of approximately $8.6\%$.} This precision will further improve with the inclusion of more events as $\sqrt{N}$. In the future, this measurement will shed light on an unexplored domain of fundamental physics and will bring deeper insights into the phenomenon of cosmic acceleration. The operation of the gravitational wave detector in the deci-Hertz frequency band is key to exploring this frontier of fundamental physics. 
	
}


\begin{document}
	
	\maketitle
	\flushbottom
	
	\section{Introduction}


One of the many tests of general relativity (GR) is to check for the speed of propagation of gravitational waves (GWs) \citep{Nishizawa2018}. Since the detection of the first GW \citep{116.061102} by the LIGO (Laser Interferometer Gravitational-Wave Observatory)-Virgo-KAGRA (Kamioka Gravitational Wave Detector) Collaboration \citep{Aasi2015, Acernese2015, 2019}, we have detected multiple events which have broadened our understanding of the Universe in several areas of astrophysics\citep{Berti:2022wzk, Adhikari:2022sve}. The first GW multi-messenger event GW170817 has also brought deeper insight into the alternative theories of the gravity by putting tightest constraints on the difference in the speed of GW (denoted by $c_{gw}$) from the speed of light as $ -3\times 10^{-15} \leq \frac{c_{gw}}{c} -1 \leq 7 \times 10^{-16}$ \citep{PhysRevLett.119.161101}. Though such constraints bring deeper insight into fundamental physics \citep{LIGOScientific:2017zic,DMemuslsion,Mukherjee:2020mha, Afroz:2023ndy, Afroz:2024oui, Afroz:2024joi, Afroz:2024lou}, understanding the nature of dark energy and the effect of any cutoff scale of the Effective Field Theory (EFT) for dark energy on the GW propagation still remains unknown \citep{deRham2018}. The above constraint has ruled out many alternate GR theories\citep{TGRGW170817}, but this constraint applies only to GW frequency bands in the LIGO band($10$ \si{\hertz} to $\sim 10^2$  \si{\hertz}). As a result, these bounds are not expected to be valid at lower frequencies of the GW signal. 

The EFT description of dark energy, which explains cosmic acceleration at the largest Gpc scales, can in principle depend on energy scale, resulting in a frequency-dependent propagation speed of GW $c_{gw}(f)$.  Such energy scale dependence arises from the EFT cutoff scale of dark energy (denoted by $M$) \citep{deRham2018, horndeski2022secondorder}. The presence of a new energy scale will have an observable impact on the GW signal due to the difference in the propagation speed of the GW signal at different frequencies, leading to a dispersive effect. Though measuring the signature of this effect from only GW observations from a narrow-frequency window is difficult, multi-band GW observations spanning 5-6 decades in frequencies are a direct probe to this effect. As a result, by combining observations from the hecto-Hertz band (such as LIGO, Virgo, KAGRA, LIGO-Aundha\citep{UNNIKRISHNAN2013}, Cosmic Explorer\citep{reitze2019cosmic} and Einstein Telescope\citep{Punturo:2010zz}) with the milli-Hertz (such as LISA (Laser Interferometer Space Antenna) \citep{PhysRevD.93.024003}) and deci-Hertz detector (such as LGWA (Lunar Gravitational Wave Antenna) \citep{Harms2021, Ajith:2024mie}) we can explore the difference in the GW dispersion relation at different frequencies and discover (or rule out) the associated energy scale of EFT description of dark energy. 



\begin{figure}
	\includegraphics[width=15cm]{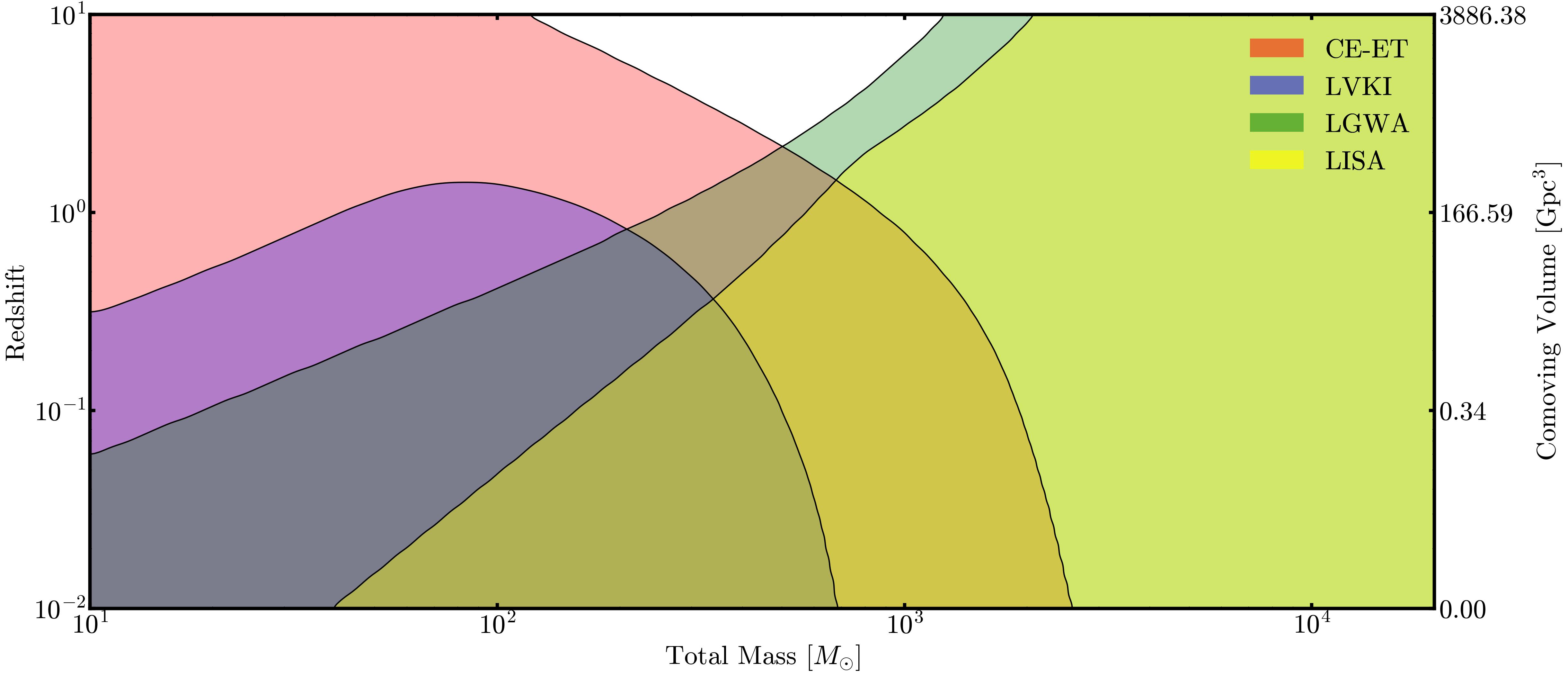}
	\caption{Sensitivity of various detectors with respect to total mass in source frame and redshift (Mass ratio $q = 1$ here). For LVKI and CE-ET, the IMRPhenomHM model, and for LISA and LGWA, the TaylorF1 model \citep{Ajith} was used for creating this plot. For the analysis of this paper, only IMRPhenomHM model is used as stated in section \ref{Method of Analysis}.}
	\label{fig_sens}
\end{figure}

Such multi-band GW observation is primarily possible from those sources which can be well-detected with a high matched-filtering signal to noise ratio (SNR) in different bands. Also, to increase the chance of detecting sources, we need to detect sources up to a high redshift covering a large cosmic volume. In figure \ref{fig_sens} we show the horizon of detecting sources for different GW detectors with an SNR above ten as a function of their source-frame total mass and redshift, along with the comoving cosmic volume up to that redshift. It shows that GW sources with total mass between about a hundred solar mass to a few thousand solar mass, which are usually referred to as intermediate mass black holes(IMBHs), are detectable up to a redshift $z\sim 1$, covering a significantly large fraction of comoving volume than possible from the sources outside this range. As a result, these are the most promising GW sources that can help study the EFT description of dark energy using multi-band observations. The multi-band observation with LISA and LVKI for binary black holes (BBHs) of masses similar to GW150914 were explored previously \citep{Baker2023,Harry2022}. 

\begin{figure}
	\includegraphics[width=15cm]{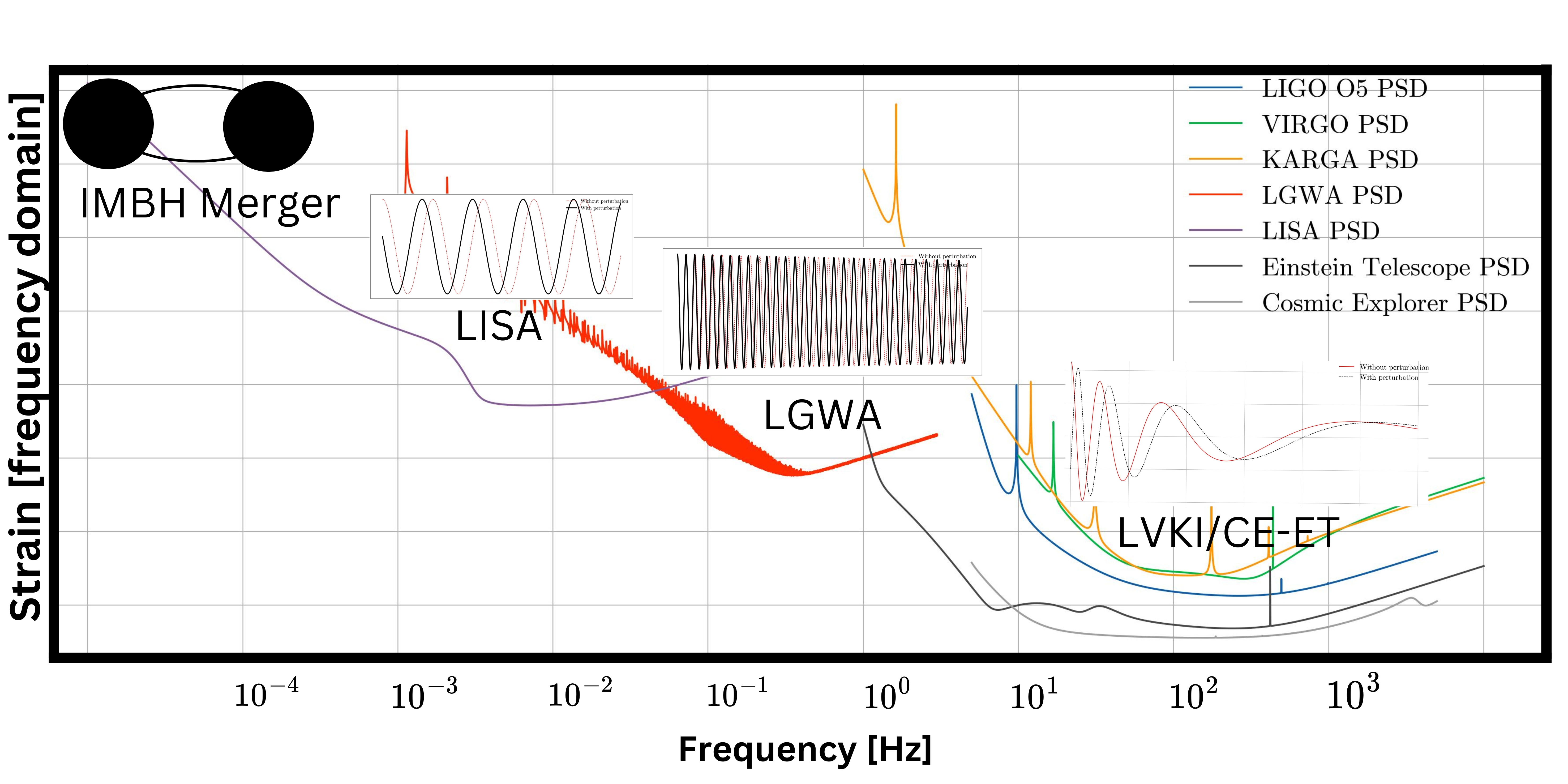}
	\caption{This figure shows the signature of dispersion due to the impact of the cutoff scale of any EFT of dark energy on the GW signal from a single IMBH source, which is detectable at different GW frequency bands, such as in the milli-Hertz range (using LISA), deci-Hertz range (using LGWA) and hecto-Hertz range (using LVKI and CE/ET).}
	\label{fig_summary}
\end{figure}

The focus on intermediate-mass black holes (IMBHs) is driven by their detectability with the detectors considered as shown in figure \ref{fig_sens}, as well as conducting analysis within the operational lifetime of the detectors. An active galactic nuclei (AGN)-motivated population model is utilized, as hierarchical mergers in AGN may serve as viable sources for IMBHs\citep{PhysRevD.105.063006,ArcaSedda2021}. Rather than relying on a single source, analyzing a population of sources enhances the robustness of findings and facilitates the investigation of redshift dependence in dispersion effects. The estimation of length scales at which the non-GR effects become significant in the analysis of the cause of deviation from GR.
Incorporating multiple detectors enables a thorough examination of the length scales at which non-GR effects might emerge, as illustrated in figure \ref{fig_summary}.

This paper is organized as follows. Section \ref{Theoretical framework of analysis} discusses the theoretical framework behind the EFT of dark energy. In section \ref{Population Model of Black Holes}  and 
section \ref{Method of Analysis} we discuss the 
underlying population of IMBHs used in the analysis and describe the analysis technique respectively. The results are discussed in Section \ref{Results} and the main conclusions and future prospects are discussed in \ref{Conclusions}. Throughout the paper, we follow natural units unless they are specifically mentioned.

\begin{figure}
	
	\includegraphics[width=15cm]{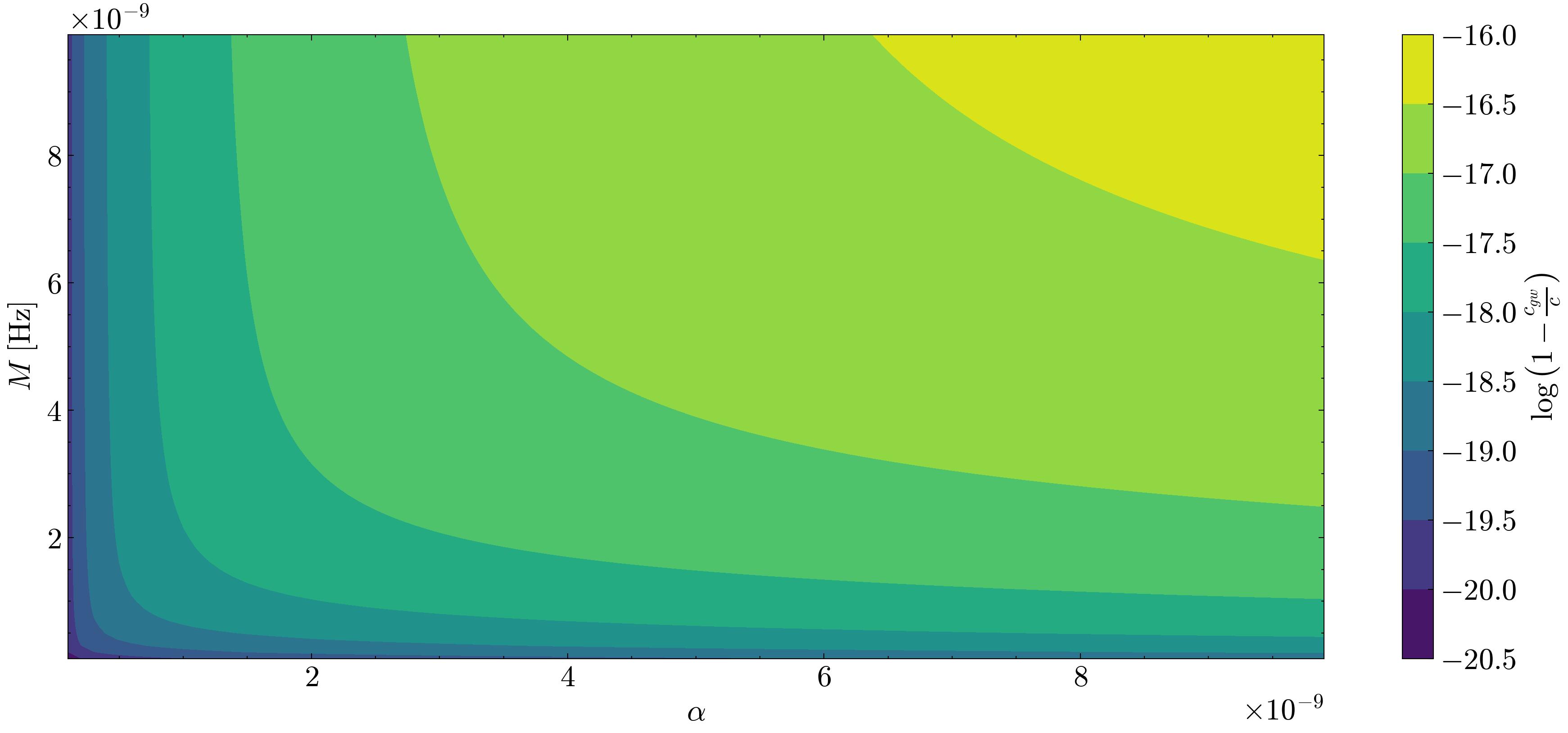
	}
	\caption{
     Strength of the dispersive effect at $f= 0.26$ \si{\hertz} for different choice of $\alpha$ and $M$. Here transition frequency $f_t$ varies with $\alpha$ and $M$. See equation \eqref{eq:M} for the relation between $\alpha$, $M$ and $f_t$.}
	\label{fig2.1}
\end{figure}

\section{Theoretical formalism: Signature of EFT of dark energy on GW}
\label{Theoretical framework of analysis}
 {The EFT of dark energy can capture a wide range of modified gravity and proposes observable signatures on different cosmological probes (see these review articles for more details \citep{Gubitosi:2012hu, Frusciante:2019xia}). Our work is focusing on the feasibility of inferring the propagation speed of GW to study the modified gravity theories.}  
In this section discusses the dispersion relation and the corresponding non-GR effects in the GW waveform. 
The propagation of GWs can be written as 

\begin{equation}
	h_p'' + 2\mathcal{H}h_p' +  k^2 h_p = 0,
	\label{eq3}
\end{equation}
where $h_p$ denotes the GW strain in p polarization ($p \in \{+,\times\}$),  {the primes denote derivatives with respect to conformal time}, $\mathcal{H}$ is Hubble parameter in conformal time,  {$c$ is the speed of light and $k$ is the wavenumber}.
However, in alternate theories of gravity \eqref{eq3} is modified as \citep{deRham2018}

\begin{equation}
	h_p'' + 2(1-\gamma(z))\mathcal{H}h_p' + (c_{gw}^2k^2 + m_{gw}^2 a^2)h_p = a^2\Pi_p,
	\label{eq4}
\end{equation}
where $\gamma(z)$ is the frictional term, $c_{gw}$ is the speed of GWs,  {$a$ is the scale factor}, $m_{gw}$ is the mass of graviton, and $\Pi_p$ is the anisotropic stress term. The effect of a non-zero frictional term is to reduce the amplitude of the strain and does not affect the time of the merger. Also, due to the current bounds of LIGO (especially from GW170817), the effect of non-zero graviton mass is negligible in the frequency bands under consideration \citep{LIGOScientific:2017zic}.  {In our analysis, we have taken the case where $\gamma(z) = m_{gw} = \Pi_{p} = 0$}. In this scenario, the GW strain in the frequency domain is given by,
\begin{equation}
	\Tilde{h}_p^{NG} = \Tilde{h}_p^{GR}\hspace{1mm} e^{-ik\Delta T},
	\label{eq5}
\end{equation}
where $\Tilde{h}_p^{GR}$ is the GW waveform from GR, $\Tilde{h}_p^{NG}$ is the GW waveform with non-GR perturbations and $\Delta T$ is given by
 {
\begin{equation}
	\Delta T = \int_0^z \frac{\delta_g d\tilde{z}}{(1+\tilde{z})\mathcal{H}},
	\label{eq6}
\end{equation}
}
where $z$ is redshift and \( \delta_g = 1-c_{gw}\).

For $\delta_g$, we refer to a phenomenological scenario as described in \citep{deRham2018} in which authors have used Horndeski effective field theory to get an analytical expression. Consider a scalar field $\phi$, represented by an effective field theory with cutoff scale $M$. On addition of correction terms at a scale $M$($M \ll \Lambda$) we get a Lagrangian for the scalar field $\phi$ as
\begin{equation}
	\mathcal{L} = -\frac{1}{2}(\partial\phi)^2 + \frac{1}{2\Lambda^4}(\partial\phi)^2\frac{M^2}{M^2 - \Box}(\partial\phi)^2,
	\label{eq7}
\end{equation}
where $\Box$ is the D'Alembertian and we define $\delta\phi = \phi - \langle\phi\rangle$ as fluctuations on top of the background given by $\langle\phi\rangle = \alpha\Lambda^2 t$. Here $\Lambda$ is the strong coupling scale, and $M$ would be referred to the energy scale of the effective theory of modified gravity scenarios such as the UV cutoff scale for the Horndeski theories \citep{deRham2018}, which can also point to the energy scale of the coupling between dark energy and gravity. From this, we get the dispersion relation for the fluctuations $\delta \phi$ as,
\begin{equation}
	\omega^2 = k^2 - \left(\frac{4\alpha^2}{1+2\alpha^2} \right)\left(\frac{M^2\omega^2}{M^2 - \omega^2 + k^2}\right),
	\label{eq8}
\end{equation}
where $\omega$ is angular frequency and $k$ is wavenumber. On simplification, we get

\begin{equation}
	\frac{1}{c_{gw}^2}-1
	= \frac{-M^2 + 
		\sqrt{M^4 + \frac{16\alpha^2 M^2 \omega^2}{1 + 2\alpha^2}}
	}{2\omega^2},
	\label{eqcgw}
\end{equation}
where $c_{gw}$ is the speed of GW. In this analysis, as the goal is to find the least deviation from GR we can detect, the approximation of $\alpha \ll 1$ can be taken($\alpha < 10^{-3}$). Figure \ref{fig2.1} shows the non-GR effect we are looking for a range of values for the parameters $\alpha$ and $M$.  In this limit, at frequencies $f \ll f_t$, the deviation is approximately,
\begin{equation}
	\delta_{g} = 1 - c_{gw} = 2\alpha^2.
	\label{eqdg}
\end{equation}
A detailed calculation of the EFT in the low frequency limit is given in appendix \ref{Effective field theory calculations}.

\begin{figure}
	\centering
	\includegraphics[width=15cm]{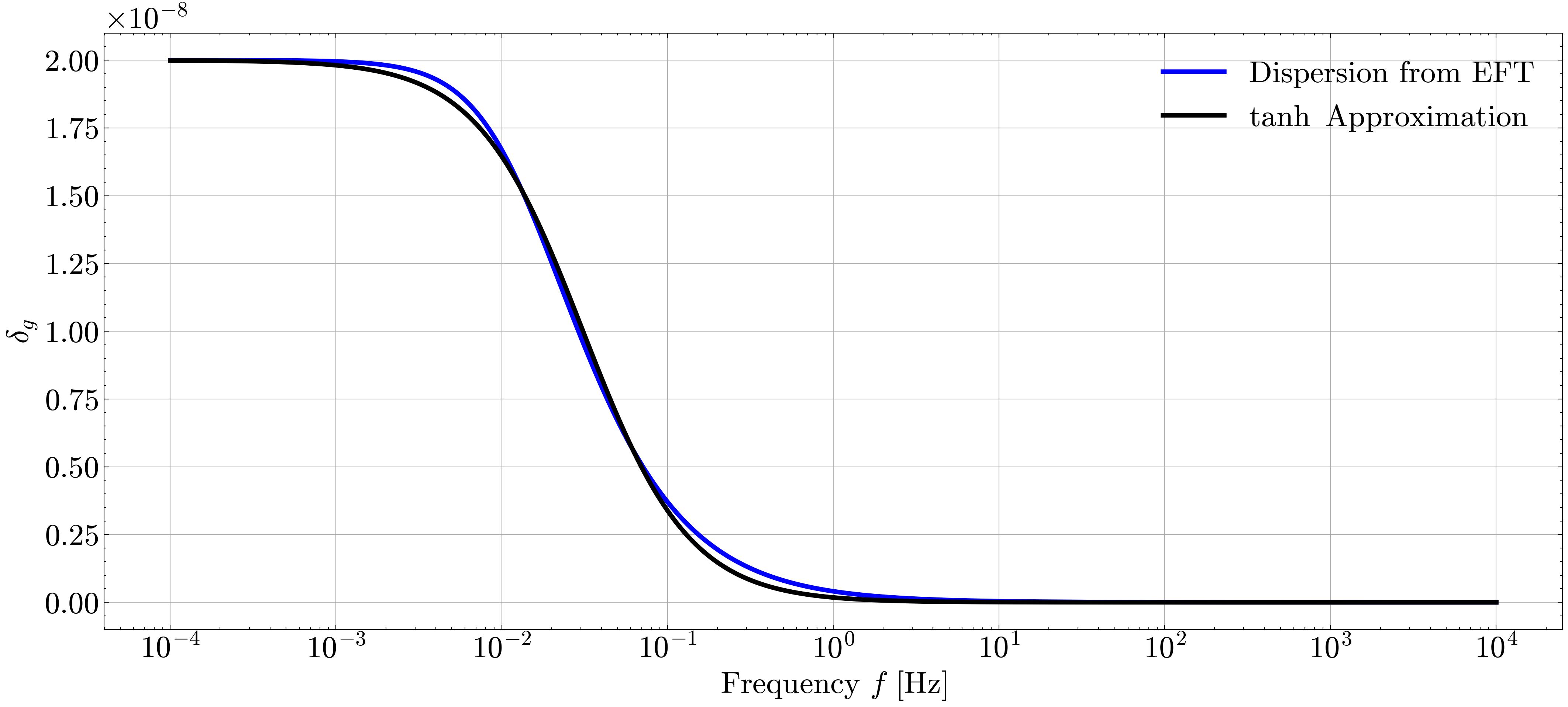}
	\caption{Dispersion $\delta_g$ vs frequency for $\alpha = 10^{-4}$ and $f_t = 0.26$ \si{\hertz}, along with the tanh approximation of $\delta_g$ at these values.  {The value of $\alpha$ is taken as $10^{-4}$ instead of $10^{-9}$, as for the latter value, equation \eqref{eqcgw} gives numerical error while performing the analysis in python}.}
	\label{figap}
\end{figure}
The numerical calculation of equation \eqref{eqcgw} for small values of $\alpha$ exhibits numerical errors in Python. To mitigate this issue, we have approximated equation \eqref{eqcgw} in the $\alpha \ll 1$ limit($\alpha <  10^{-5}$) with a $\tanh$ function given by
\begin{equation}
	\delta_{g} = \alpha^2[1 - \tanh(g_1 s (\log_{10}(2\pi f) - \log_{10}(2\pi f_t)) + g_2)],
	\label{eq_main_approx}
\end{equation}
where $g_1 = 0.6871$ and $g_2 = 0.1968$ are estimated from fitting and the frequency $f_t$ is defined as transition frequency and is given as,
\begin{equation}
	f_t = \frac{M\sqrt{16 + 4\sqrt{17}}}{8\pi\sqrt{2}\alpha}.
	\label{eq:M}
\end{equation}
The parameter $s$ in equation \eqref{eq_main_approx} is given as,
\begin{align}
	s &= \exp\left(\frac{N_1-N_2}{16\pi^2 f_t^2 - M^2\beta}\right),  \\ \nonumber
	N_1 &= \frac{16 \pi^2 f_t^2 - 2\pi M^2 \beta_d f_t }{\sqrt{2}(8\pi^2 f_t^2 - M^2\beta)}, \\ \nonumber
	N_2 &= \sqrt{16\pi^2 f_t^2 - 2 M^2 \beta}, \\ \nonumber
	\beta_d &= -\frac{32\pi \alpha^2 f_t}{\beta} ,\\  \nonumber
	\beta &= 1 - \sqrt{1 + \frac{64 \alpha^2 \pi^2 f_t^2}{M^2}}.
\end{align}
Figure \ref{figap} compares the $\tanh$ approximation and the actual curve. The approximation fits really well with the actual curve except for slight deviations near the transition frequency $f_t$. However, since the deviations can be considered negligible with respect to noise in our analysis, this approximation can be considered valid for doing the study. 

\begin{figure}
	\centering
	\includegraphics[width=17cm]{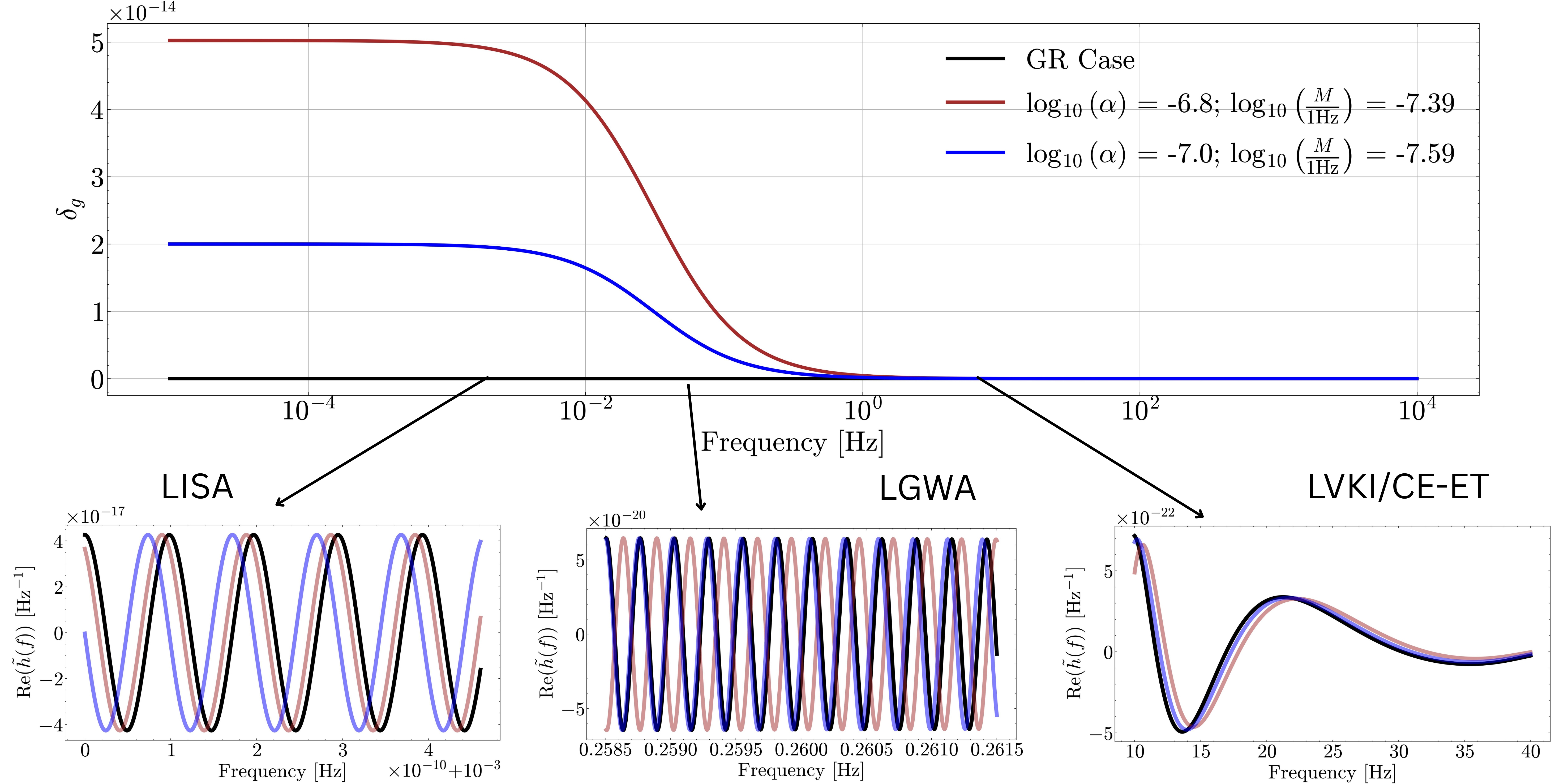}
	\caption{Deviation in speed of GW $\delta_{g}$ vs frequency for multiple $\alpha$ and $M$ values for a transition frequency of $0.26$ \si{\hertz}. The waveform plots show the effects of this perturbation for different $\alpha$. The source is a 200 $M_{\odot}$ and 200 $M_{\odot}$ BBH merger. The black lines in the waveform plots indicate the GW waveform in the absence of any non-GR effects.}
	\label{fig:dev}
\end{figure}

The impact of this dispersion signal on different frequency range of GW signal is shown in figure \ref{fig:dev} for two different strengths of the parameters $\alpha$ and M in comparison to the GR case shown in black. At frequencies below the transition frequency $f_t$, a constant shift in GW signal happens in the milli-Hertz band due to a non-zero difference of the GW propagation speed from the speed of light. However, at the frequency range which are close to the transition frequency $f_t$, the shift in the GW signal in the deci-Hertz band varies with frequency as shown in the middle panel. However, in the high frequency range, the difference in the GW signal in comparison to the GR scenario is less pronounced than in the low frequency regime. The combination of the signal in all these three frequency ranges allows the capture of the strength of the dispersion and also the transition frequency, which provides a direct probe to the energy scale of EFT (as shown in equations \eqref{eqdg} and \eqref{eq:M}).



\section{Population Model of Binary Black Holes}
\label{Population Model of Black Holes}

This section describes the mass and redshift distribution model of IMBHs, motivated by AGN distribution, used in the analysis. As we are interested in the IMBH population of BBHs in the Universe, we have considered the AGN discs as a possible formation scenario of the IMBH mergers, motivated by several previous studies \citep{Yang2019, Yang2020, 2023MNRAS_}. 

\subsection{Mass distribution of black holes}
The mass range of the black holes considered lies in the intermediate mass region. This range is detectable within the various detector bands and the noise characteristics as shown in figure \ref{fig_sens}. Moreover, IMBHs evolve from the LISA band to the LIGO band with a timescale in the order of a year, and hence, with these sources, we can perform multi-band analysis within the lifetime of the detectors. Furthermore, IMBHs can also be detected up to higher redshift due to stronger signal, which helps us increase the number of detections a lot more than by using stellar mass black holes.

\begin{figure}[H]
	\centering
	\includegraphics[width=17cm]{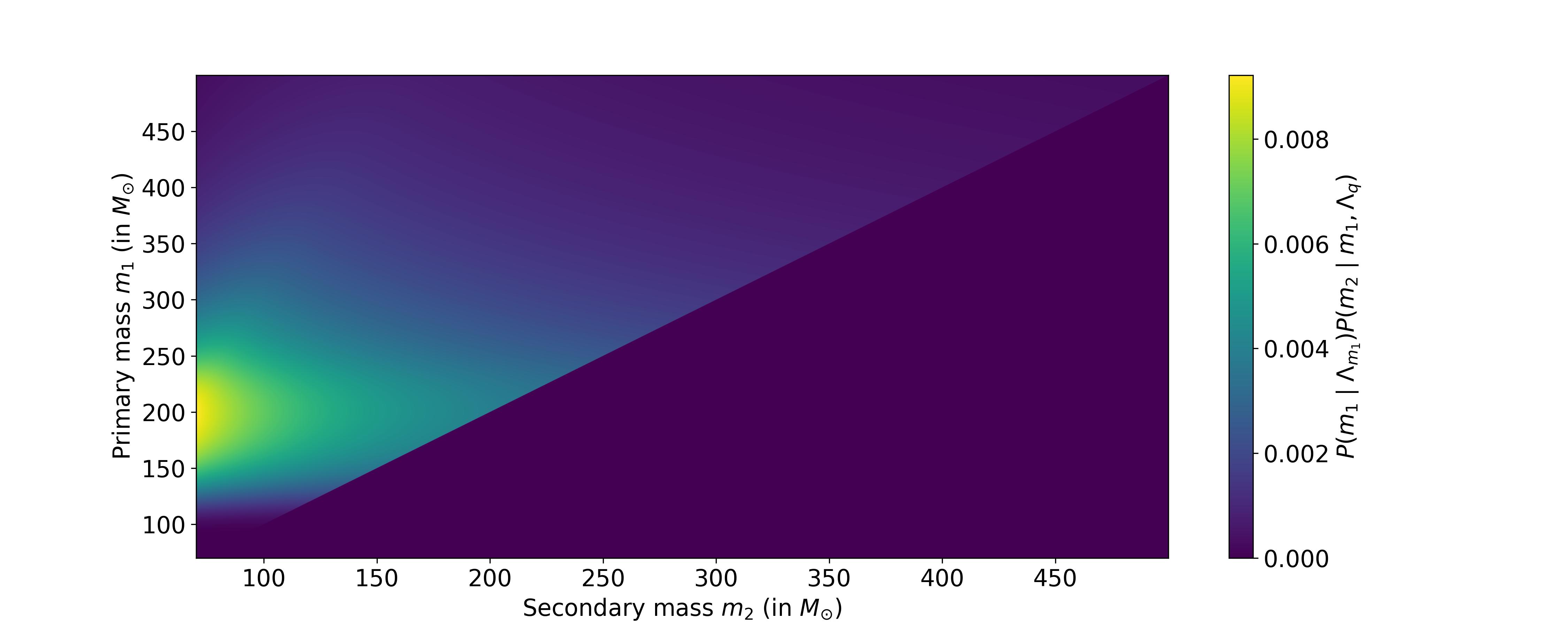}
	\caption{Joint probability distribution of primary and secondary masses of IMBHs considered in this analysis. More details on the modeling is given in appendix \ref{Mass model}.}
	\label{figq}
\end{figure}

We consider a smoothened power law similar to \citep{Yang2019, Yang2020,2023MNRAS_,Karathanasis2023} is used for the mass distribution, an extension of the mass distribution for stellar mass black holes from LIGO and information from hierarchical mergers. A detailed description is given in appendix \ref{Mass model}. The joint distribution of BBH population for both the component mass are shown in figure \ref{figq} with probability in color-bar. 

\subsection{Redshift distribution}
\label{Redshift distribution}

\begin{figure}
	\centering
	\includegraphics[width=17cm]{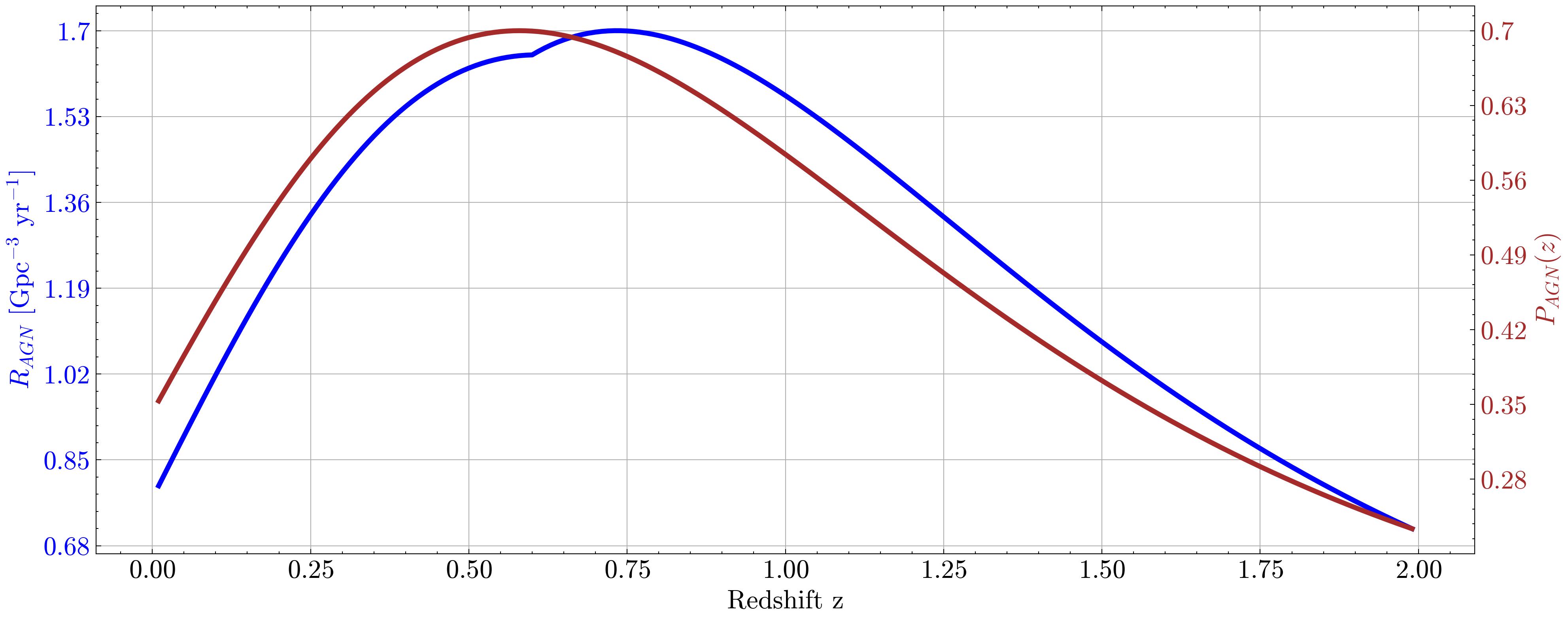}
	\caption{Distribution of AGN population, $P_{AGN}$ and merger rate of IMBHs, $R_{AGN}$ vs redshift.  {The blue line denotes $R_{AGN}(z)$ while the brown line denotes $P_{AGN}(z)$}.}
	\label{fig:containsfig_red}
\end{figure}

An AGN motivated population model is used for the redshift distribution because AGN disks could be a location of the population of IMBHs. These IMBHs are formed from hierarchical mergers starting from stellar-origin black holes \citep{PhysRevLett.123.181101}. Thus, the merger rate and population density of IMBHs considered here are correlated with the AGN population distribution. Since we are using IMBHs which can be detected farther away, we are considering sources up to a redshift of two. This will enable us to identify more sources, as higher redshifts correlate with a larger volume and an increased prevalence of AGNs. The detector sensitivities shown in figure \ref{fig_sens} indicate that detecting sources up to these redshifts for the mass range considered is possible. For a detailed discussion of the distribution of IMBHs with redshift, the reader is referred to appendix \ref{Redshift model}. Figure \ref{fig:containsfig_red} illustrates the merger rate of IMBHs and the distribution of AGN populations with respect to redshift considered in this analysis following the previous IMBHs formation scenario in AGNs \citep{Yang2020}. The plot in blue shows the merger rate distribution of BBHs and in red the AGN distribution is shown. The shift between the blue and red curve arises due to the model of the redshift evolution of the Eddington ratio distributions and the dependence of the BBH merger on the accretion rate (see the paper \citep{Yang2020} for more details). It is important to mention that though the redshift evolution of the BBHs depends on the model considered, our main scientific result on the constraints on the individual parameters primarily depends on the high SNR events and the error on the estimation of their  {arrival time (as discussed in the next section)}, and less susceptible to the exact distribution of the BBH. In future, with the availability of more data, a better IMBHs population model can be considered for the joint multi-band analysis. 

To simulate the population of BBH mergers the following formula for total number of mergers is used,
\begin{equation}
	\frac{\text{d}N_{gw}}{\text{d}z} = \frac{R_{AGN}(z)}{1+z}\frac{\text{d}V_c}{\text{d}z} T_{obs},
	\label{eq17}
\end{equation}
where $R_{AGN}$ is the AGN motivated merger rate described in section \ref{Redshift distribution}, $V_c$ is the comoving volume and $T_{obs}$ is the time period of observation. We consider events occurring in $z \in (0.01,2)$ for $T_{obs} = 1\text{ year}$ and hence get $N_{gw} = 371$. The masses and redshifts of these sources are randomly selected from the mass distribution and redshift distribution described above. 


\section{Method of GW Analysis}
\label{Method of Analysis}
\begin{figure}
	\centering
	\includegraphics[width=15cm]{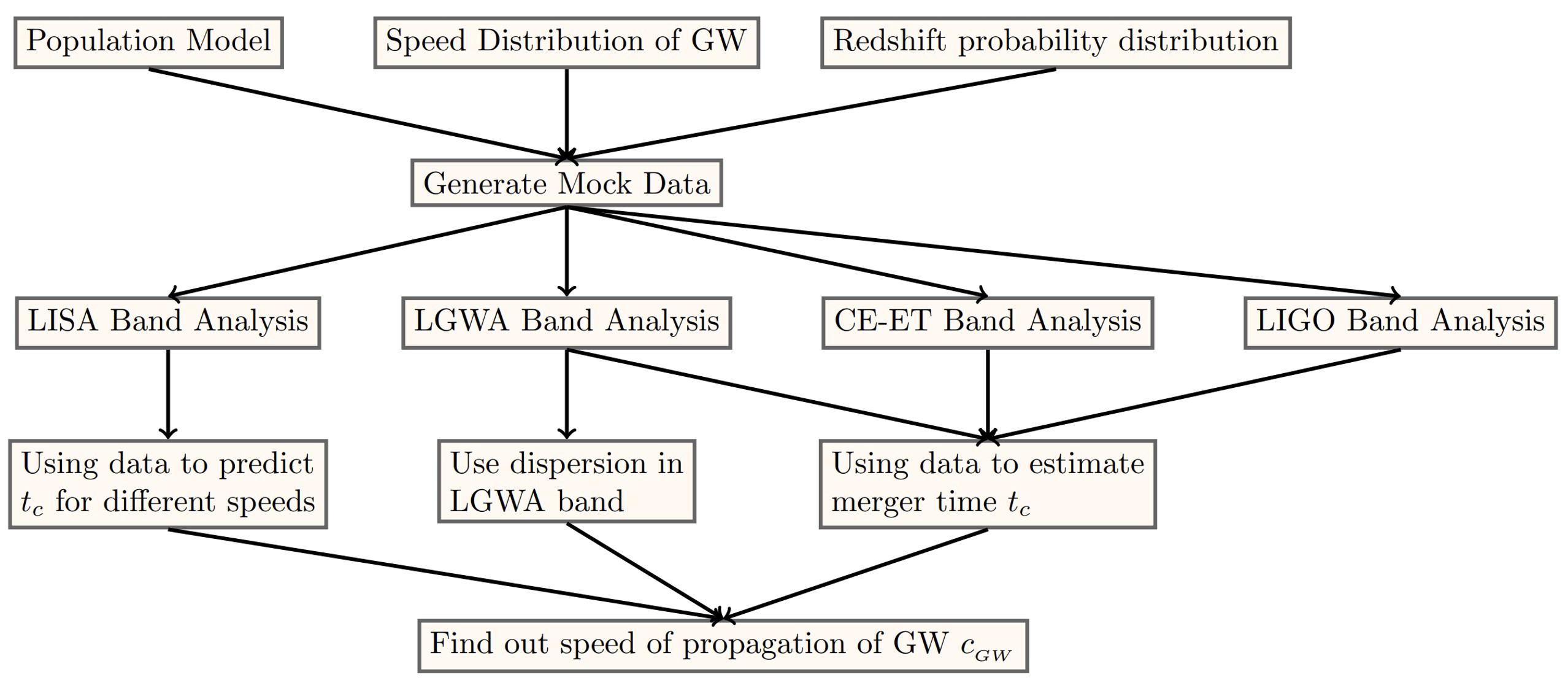}
	\caption{Flow chart of the analysis}
	\label{fig:figflow}
\end{figure}

In this section we discuss the generation of mock samples and the data analysis method used, along with a brief review of GW signal. We describe the flowchart of the analysis in figure \ref{fig:figflow}. 

\subsection{GW signal}
A GW signal is a tensor composed of two degrees of freedom corresponding to the plus and cross polarizations respectively. 

Consider an arbitrary GW traveling in the $\hat{k}$ direction. We can then find two vectors orthogonal to $\hat{k}$, which are also mutually orthogonal, say $\hat{u}$ and $\hat{v}$. These three vectors form an orthonormal basis. To describe the GW, which is a tensor, basis tensors are required, which are defined as,
\begin{align}
	\mathbf{e}^{+} = \hat{u}\otimes\hat{u} - \hat{v}\otimes\hat{v},\\
	\mathbf{e}^{\times} = \hat{u}\otimes\hat{v} + \hat{v}\otimes\hat{u}.
\end{align}
Using this, an arbitrary GW signal can be represented as 
\begin{equation}
	\mathbf{\tilde{h}}(f) = \tilde{h}_{+}(f) \boldsymbol{\epsilon}^{+} + \tilde{h}_{\times}(f) \boldsymbol{\epsilon}^{\times},
\end{equation}
where $\tilde{h}_{+}(f)$ and $\tilde{h}_{\times}(f)$ are the plus and cross-polarization frequency domain strain respectively, and,
\begin{align}
	\boldsymbol{\epsilon}^{+} = \cos(2\psi)\mathbf{e}^{+}- \sin(2\psi)\mathbf{e}^{\times}, \\
	\boldsymbol{\epsilon}^{\times} = \sin(2\psi)\mathbf{e}^{+}+ \cos(2\psi)\mathbf{e}^{\times}.
\end{align}
Here $\psi$ is the polarization angle. A measured GW signal can in general be written as 
\begin{equation}
	\tilde{h}(f)_{det} = \mathbf{D}(f,t,\hat{k}):\mathbf{\tilde{h}}(f).
	\label{eq:det_h}
\end{equation}
 {Here} $\mathbf{D}(f,t,\hat{k})$ is the detector tensor at frequency $f$, time $t$ and direction of propagation $\hat{k}$ and  {"$:$" denotes tensor contraction}. Equation \eqref{eq:det_h} can be expressed in a scalar format as,
\begin{equation}
	\tilde{h}(f)_{det} = F_{+}(f,t,\hat{k}) \tilde{h}(f)_{+} + F_{\times}(f,t,\hat{k}) \tilde{h}(f)_{\times},
\end{equation}
where $F_{+}$ and $F_{\times}$ are called the antenna parameters.

This GW strain from a BBH mainly consists of three phases, the inspiral, merger and ringdown phase. Since we are considering IMBHs, the inspiral phase will lie in LISA and LGWA detectors, whereas the merger and ringdown phases will lie in CE-ET and LVKI detector. The generation of this gravitational wave strain is done by using \textbf{IMRPhenomHM} \citep{Garca2020} model. This model includes the higher modes of GWs, which would be important in the CE-ET/LVKI band as there are only a few cycles of the GW in this band, and the effect of higher modes cannot be neglected. The higher modes in the waveform will help us resolve the degeneracy between luminosity distance and inclination angle and test the waveform for non-GR effects more accurately. For this analysis, only non-spinning and non-precessing black holes are considered. \textbf{Bilby}\citep{Ashton2019} is used for generating the GW waveform for the ground based detectors, and also to get the detector response of the same. For LGWA, \textbf{GWFish}\citep{Dupletsa2023} is used to get the antenna response characteristics. The waveform generation and antenna parameter calculation for LISA are performed using \textbf{BBHx} \citep{michaelkatz2021,PhysRevD.102.023033,PhysRevD.105.044055}. 

The matched filtering SNR of the signal denoted by $\rho$ is calculated as
\begin{equation}
	\rho^2 \equiv 4 \text{ Re} \int_{f_{min}}^{f_{max}} \frac{\tilde{h}(f)|^2}{S_n(f)} \mathbf{d}f,
	\label{eq20}
\end{equation}
where $S_n(f)$ is the power spectral density. The SNR bounds are taken as 10 in LVKI, 50 in CE-ET, 10 in LGWA. We want our signal to be detected in either CE-ET or LVKI, as well as be detected in LGWA. Since there will be a detection in LGWA, we did not put any bound in LISA, as with LGWA itself we can measure dispersion. Based on these bounds we have calculated that there are seventeen such golden events in a period of one year. It should be noted that even though we did not put a bound in LISA, there we five sources with SNR 10 and above in LISA from the seventeen sources considered. Also, the seventeen sources are a result of a random realization of the mass and redshift distribution and can change for different runs, due to Poisson fluctuation. In the remaining analysis we will use these seventeen golden events for the forecast of measuring the energy scale of EFT of dark energy. 

\subsection{Measuring the uncertainty on GW source parameters}
On the seventeen golden events, fisher matrix analysis is performed to compute the error bars for the non-GR parameters. The parameters of interest for each source are the amount of dispersion which is quantified by $\alpha$, transition frequency $f_t$ and time of merger $t_c$.

To reduce the computation cost for GW source parameter estimation, we opted for Fisher matrix analysis instead of MCMC analysis to estimate the error bars associated with the parameters of interest. As currently the MCMC analysis are performed with simulated detector noise for third-generation detectors with a Gaussian covariance matrix, the Fisher analysis will be able to capture the uncertainty between the parameters and their correlation and can provide the minimum error associated with a measurement (known as the Cramer-Rao bound \citep{cramer,rao} . The elements of the Fisher matrix $\mathcal{F}_{ij}$ are computed as \citep{cramer, rao}
\begin{equation}
	\mathcal{F}_{ij} = \left(\frac{\partial \tilde{h}(f)}{\partial \theta_i} \vline\frac{\partial \tilde{h}(f)}{\partial \theta_j}\right),
\end{equation}
where $\vec{\theta} = (\mathcal{M}_c,q,d_L,\theta_{inc},\alpha,f_t,t_c)^T$, $\mathcal{M}_c$ is chirp mass, $q$ is mass ratio, $d_L$ is luminosity distance and $\theta_{inc}$is the inclination angle. The inner product is defined as
\begin{equation}
	(a\mid b) \equiv 4\text{ Re}\int_{f_{min}}^{f_{max}} \mathbf{d}f \frac{\bar{a}(f) b(f)}{S_n(f)} .
\end{equation}
Please note that the measurements of $f_t$ and $t_c$ is in SI units, and $\alpha$ is dimensionless. 

The purpose of multi-band GW analysis is probing a larger frequency range, hence giving better estimation/tight constraints on the non-GR parameter .  {The amount of dispersion can be measured by looking at the predicted time of merger from detectors such as LISA($10^{-5}$ \si{\hertz} to $10^{-3}$ \si{\hertz}) which see the inspiral phase of the IMBH, and then comparing it with the observed time of merger from detectors such as CE-ET($5$ \si{\hertz} to $10^{4}$ \si{\hertz}) and LIGO($10$ \si{\hertz} to $10^{3}$ \si{\hertz}) . LGWA ($10^{-2}$ \si{\hertz} to $1$ \si{\hertz}) is mainly used to see the transition in the dispersion effects from speed of propagation of GW signal from sub-luminal to luminal}.  {This helps in directly constraining the EFT energy scale $M$ that can be inferred from $\alpha$ and $f_t$, and hence better estimation of both parameters are necessary to get a good understanding of any violation of GR and signature of any new energy scale.} There has been other works such as \citep{Harry2022, Baker2023} on testing GR by looking for dispersion in GWs with a few tens of stellar mass sources using LVKI and LISA. Though such measurements can provide interesting hints, high precision measurement of this non-GR signal is possible from deci-Hertz GW band, in combination with ground based GW detectors such as LVKI and CE-ET. LISA with LVKI/CE-ET will provide much weaker measurement in comparison to the inclusion of a deci-Hertz detector as demonstrated in this work.




\section{Results}
\label{Results}



In this section, the results of the analysis are presented. Two scenarios were examined: (i) EFT dark energy model and (ii) fiducial $\Lambda$-CDM model, detailed below. Note that in the units we are considering, the transition frequency and energy scale are measured in the units of \si{\hertz}(see equation \eqref{eq:M}).

\begin{table}[htbp]

\centering
\small
\begin{tabularx}{\textwidth}{@{}X *{3}{S[table-format=1.3e-1]} *{3}{S[table-format=1.3e-1]}@{}}
\toprule
\textbf{Detectors} & 
\multicolumn{3}{c}{\textbf{EFT Dark Energy Model}} & 
\multicolumn{3}{c}{\textbf{Fiducial $\Lambda$-CDM Model}} \\
\cmidrule(lr){2-4} \cmidrule(lr){5-7}
& {$\sigma_{\alpha}$} & {$\sigma_{f_t}$ (\si{\hertz})} & {$\sigma_{M}$ (\si{\hertz})} 
& {$\sigma_{\alpha}$} & {$\sigma_{f_t}$ (\si{\hertz})} & {$\sigma_{M}$ (\si{\hertz})} \\
\midrule
CE-ET+LGWA +LISA    & 9.766e-12 & 2.074e-2  & 1.388e-11 & 1.609e-8  & 5.406e-2  & 5.562e-9  \\
LVKI+LGWA+ LISA  & 2.885e-10 & 2.074e-2  & 4.692e-10 & 8.922e-7  & 6.723e-2  & 3.738e-7  \\
CE-ET+LGWA      & 1.040e-11 & 2.076e-2  & 1.508e-11 & 1.048e-7  & 5.598e-2  & 3.747e-8  \\
LVKI+LGWA       & 3.257e-10 & 2.076e-2  & 5.298e-10 & 1.393e-6  & 6.725e-2  & 1.393e-6  \\
LVKI+LISA       & 1.503e-8  & 5.755e2   & 7.629e-6  & 9.976e-7  & 5.065e0   & 3.365e-5  \\
CE-ET+LISA      & 1.151e-9  & 1.975e2   & 8.962e-7  & 1.675e-8  & 9.101e-2  & 1.009e-8  \\
\bottomrule
\end{tabularx}
\caption{Error bars ($1\sigma$) for non-GR parameters in EFT dark energy and fiducial $\Lambda$-CDM models. For GR: $\alpha = 10^{-10}$, $f_t = \SI{e-7}{\hertz}$, $M = \num{6.235e-17}$ \si{\hertz}. For EFT: $\alpha = 10^{-10}$, $f_t = \SI{0.26}{\hertz}$, $M = \num{1.621e-10}$ \si{\hertz}. }
\label{tab:error bars}

\end{table}


\subsection{EFT dark energy model}
\label{Non-GR Case}

\begin{figure}
	\centering
	\hspace{-2cm}
	\begin{subfigure}[b]{0.5\textwidth}
		\centering
		\includegraphics[width=9cm]{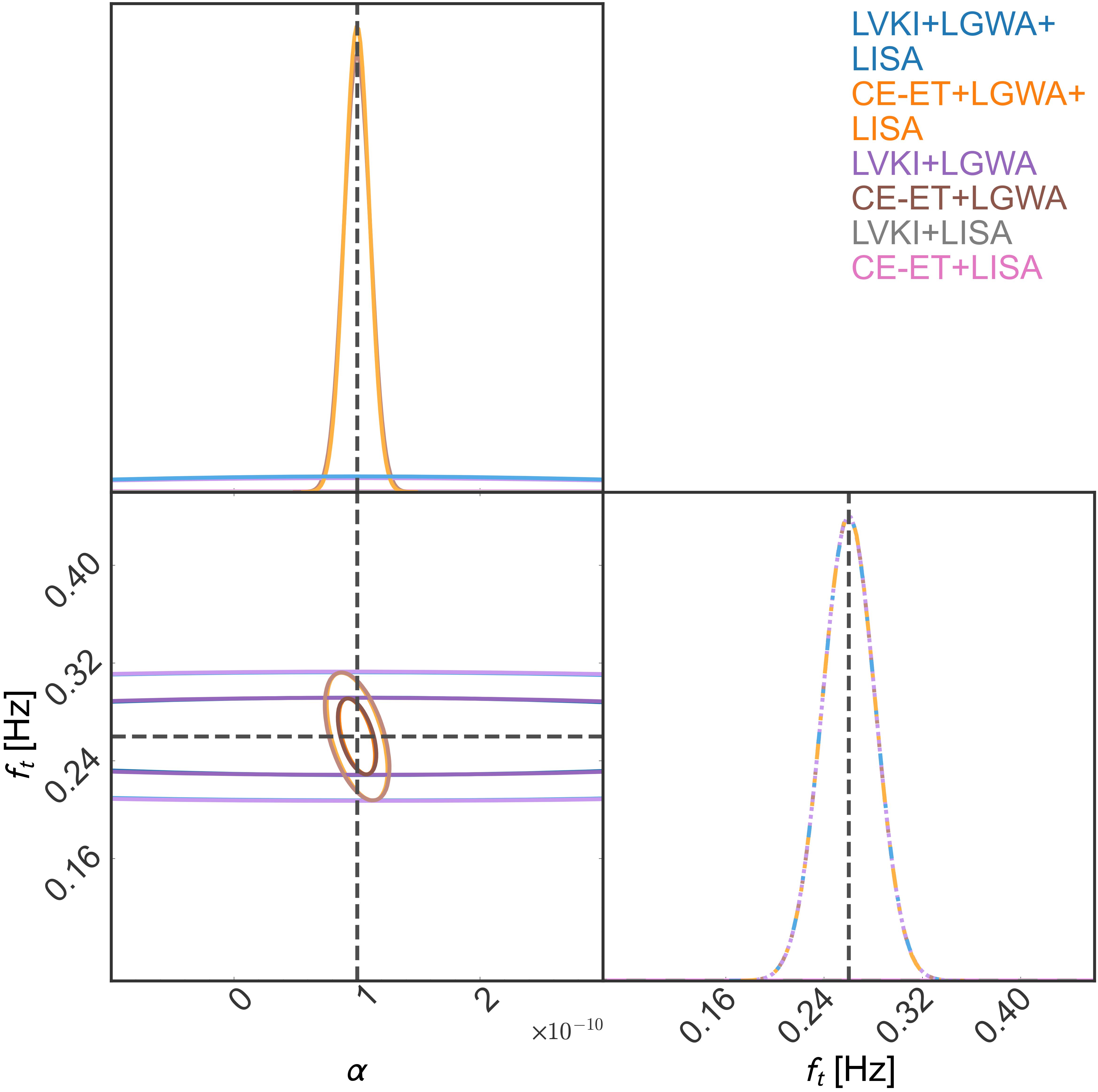}
		\caption{Posterior with $\alpha$ and $f_t$}
		\label{fig:res_NGR_ft}
	\end{subfigure}
	\hfill
	\hspace{-1cm}
	\begin{subfigure}[b]{0.5\textwidth}
		\centering
		\includegraphics[width  = 9cm]{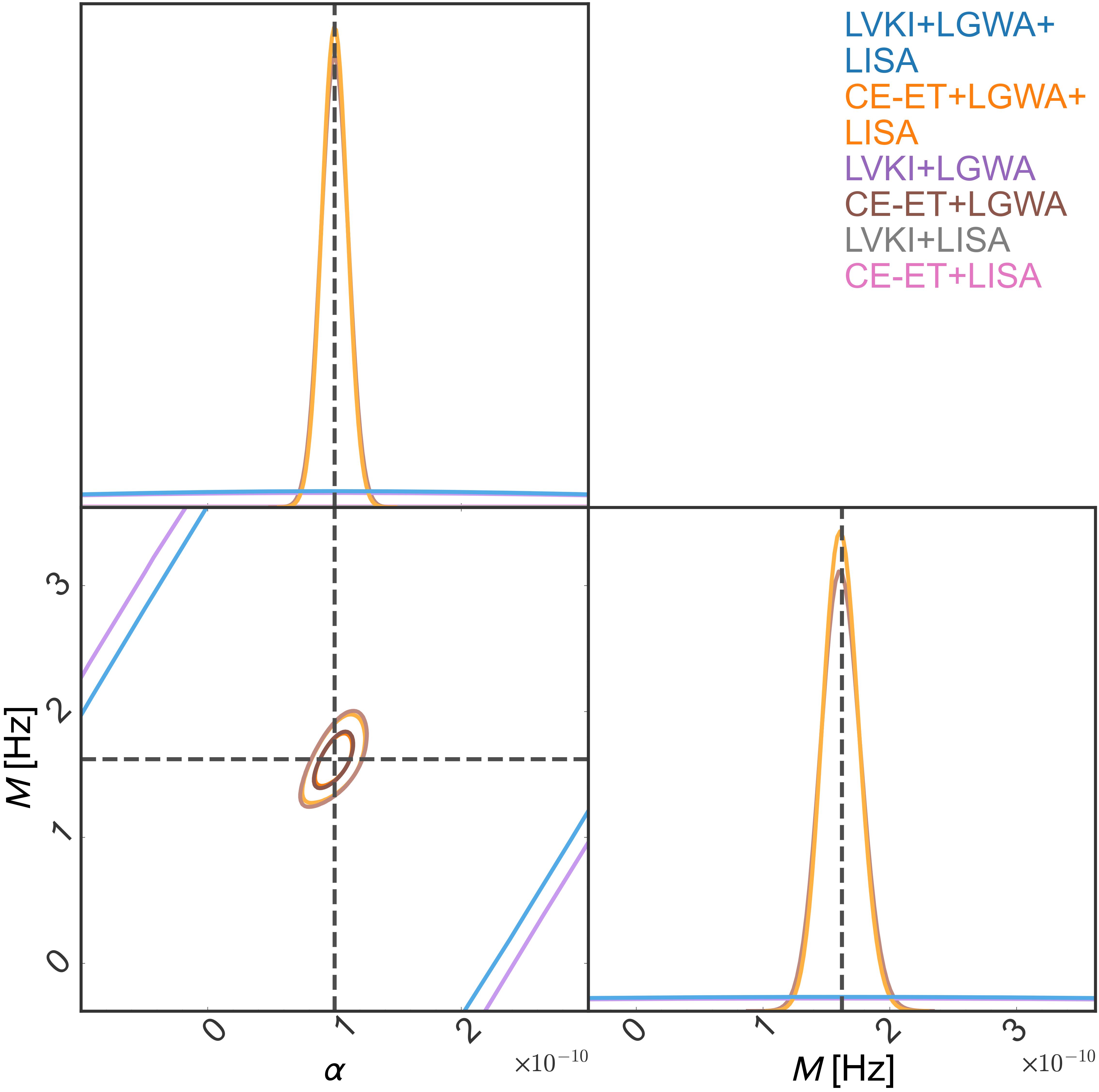}
		\caption{Posterior with $\alpha$ and $M$}
		\label{fig:res_NGR_M}
	\end{subfigure}
	\caption{The estimates on $\alpha$, $f_t$ and $M$ from various detector combinations. The estimation of $M$ is done as per equation \eqref{eq:M}. The fiducial values are $\alpha = 10^{-10}$, $f_t = 0.26$ \si{\hertz} and  {$M = 1.621 \times 10^{-10}$ \si{\hertz}}.  } 
	\label{fig:res_NGR}
\end{figure}
\begin{figure}
	\centering
	\includegraphics[width=15cm]{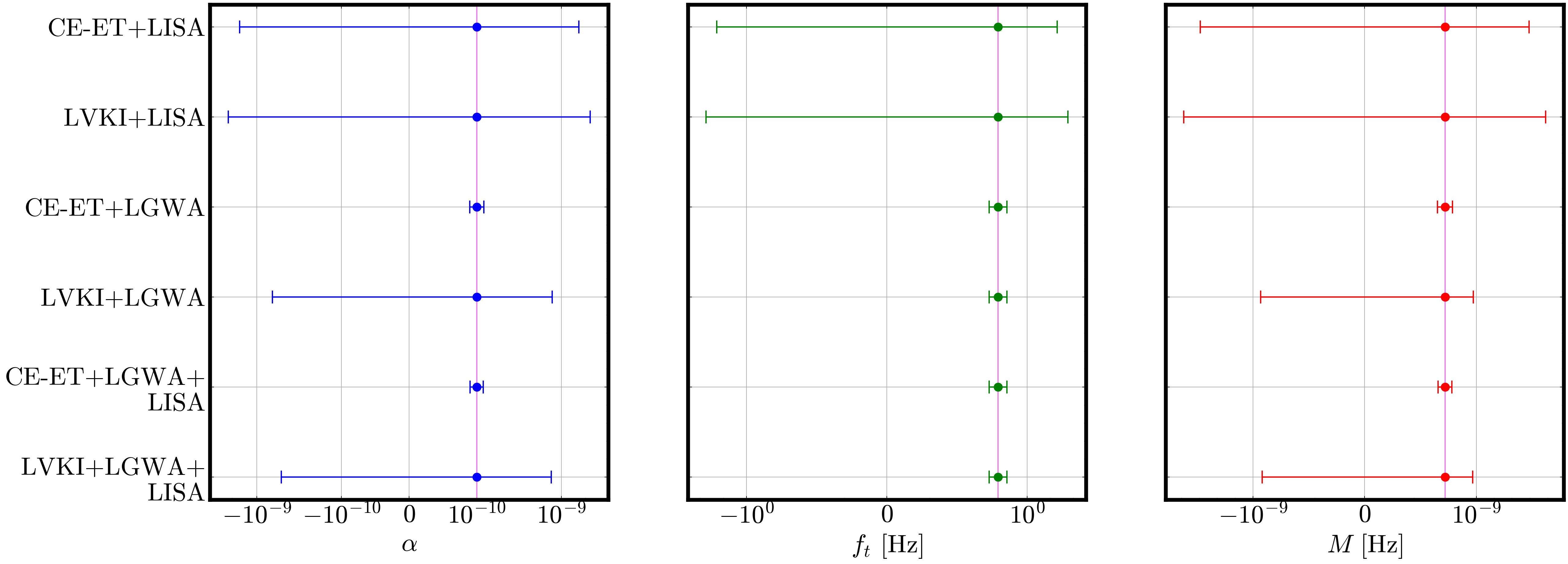}
	\caption{Error bars($1\sigma$) for various detector combinations. The LGWA combinations have better error bars than combinations without LGWA. The error bars on $M$ are estimated as per \eqref{eq_main_approx}. The x-axis is made in symmetric log.}
	\label{fig:NGR_errs}
\end{figure}

Figure \ref{fig:res_NGR} presents the estimates obtained from various detector combinations, to show those combinations which can rule out the GR scenario with 5$\sigma$ confidence level. Figure \ref{fig:NGR_errs} shows the $1\sigma$ error bars for the detector combinations considered. For ruling out the GR case, we need to rule out zero dispersion ,  $\alpha = 0$ or zero transition frequency, $f_t = 0$ \si{\hertz} with $5 \sigma$ confidence. With the fiducial values considered for the non-GR parameters and the error bars estimated for various combinations, we can eliminate the GR scenario for CE-ET+LGWA+LISA and CE-ET+LGWA as illustrated in table \ref{tab:error bars} and figure \ref{fig:NGR_errs}.

As seen from figures \ref{fig:res_NGR} and \ref{fig:NGR_errs}, measuring both $\alpha$ and $f_t$ with detectors in a single frequency band can be hard or outright not possible depending on the frequency band. Although having a detector in the frequency range with the transition frequency $f_t$ will help in estimating the observed non-GR parameters accurately, doing a multi-band analysis will help us do the same with fewer number of sources and with greater accuracy. In our analysis, the fiducial value of transition frequency lies in the deci-Hertz frequency band, and hence having a detector in this band(LGWA in our case) is crucial for getting good estimates of the non-GR parameters, especially $f_t$. Having detectors at $f \ll f_t$ (LISA in our case from $10^{-5}$ \si{\hertz} to $10^{-1}$ \si{\hertz}) and $f \gg f_t$(ground based detectors such as LVKI/CE-ET which in general has good sensitivity in $10^{1}$ to $10^{3}$ \si{\hertz}) frequency bands, at which the speed of GW is constant but differs in the two frequency regions will greatly improve the estimation of $\alpha$ which quantifies the amount of dispersion present. With just detectors in the bands of $f \ll f_t$ or $f \gg f_t$, we cannot measure dispersion as any deviation in speed of GW will give rise to a constant shift in time axis, translating to a constant phase shift in frequency domain. Hence, due to the requirement of estimating both $\alpha$ and $f_t$ accurately, to get a good estimate of energy scale of EFT, $M$, multi-band analysis is essential.  {The best estimation of $M$ is $1.621^{+0.139}_{-0.139} \times 10^{-10}$ \si{\hertz}}, feasible from CE-ET+LGWA+LISA. $M$ is a measure of the energy and length scales at which the new physics enters which causes the dispersion in GW propagation speed. Thus, presence of a non-zero transition frequency and hence a non-zero $M$ would point towards the evidence of new physics at large length scales/small energy scale . This could be due to coupling between dark energy and gravity. If so, it can lead to better understanding of cosmic expansion and a groundbreaking discovery for cosmology. A null result (or only an upper bound) would also be beneficial as this would put constraints on many alternate GR theories and eliminate many others as well. As of now, there have been no tests of GR made at these scales using GW. 

In this analysis we have fixed the transition frequency at $f_t = 0.26$ \si{\hertz}\citep{deRham2018} . This value was chosen since it is one of the highest transition frequency values than can occur which also agrees with the current constraints on speed of GW in LVKI band. However, the transition frequency can be lower than this value. The estimation of $f_t$ in our analysis is greatly improved by the presence of a deci-Hertz detector. This is partly due to fiducial value of $f_t$ lying in the deci-Hertz frequency band, and partly due to the choice of IMBHs for which deci-Hertz detectors have a high sensitivity. The choice of IMBHs is made because it evolves through the frequency bands of the various detectors we considered and this evolution occurs in the timescale of a few months to about 1-2 years, allowing us to perform multi-band analysis in a shorter duration of time and thus do this analysis within the lifetime of the detector . They also have good sensitivity in detectors in multiple frequency band as shown in figure \ref{fig_sens}, thus making this mass range of black holes a sweet spot for doing multi-band analysis. Thus, even in the fiducial value of $f_t$ is in LISA band, the presence of a deci-Hertz detector is still crucial. The importance of deci-Hertz detector for a null case is described in the next section. As long as $f_t$ stays in the LISA or LGWA frequency bands, the detector network used in this analysis can estimate $\alpha$ and $f_t$, although the number of sources and observation period required for having a precise inference can vary depending on the fiducial value. But if $f_t$ lies below the LISA frequency band, then the analysis we have done cannot be used for probing the dispersion in GW speed and we will have to opt for other techniques or detectors in other frequency band . This would essentially point towards to a null test in our analysis and is discussed in section \ref{GR Case}.

\subsection{Fiducial $\Lambda$-CDM model under GR}
\label{GR Case}

In this section, we use the analysis method mentioned in section \ref{Method of Analysis} in a scenario with no non-GR effects to show that there are no systematic errors present in our analysis and that the limiting case is satisfied. Note that with the detector network considered, non-GR effects are not completely ruled out, and the best that can be done is to put constraints on $\alpha$ and $f_t$, and hence bounds on the energy scale of EFT $M$ . Thus the limiting case we are mentioning here refers to the scenario with zero or extremely small non-GR effects which cannot be probed with the detector combinations considered. The fiducial values of the non-GR parameters for this are $\alpha = 10^{-10}$ and $f_t = 10^{-7}$ \si{\hertz}. Here, the transition frequency is below the LISA band, and hence there would effectively be no constraints on the non-GR parameters with the detectors considered. Due to this, the fiducial values chosen here are sufficient to show the limiting case of a lack of perturbation/extremely small perturbation to the GWs.

\begin{figure}
	\hspace{-2cm}
	\centering
	\begin{subfigure}[b]{0.5\textwidth}
		\centering
		\includegraphics[width=9cm]{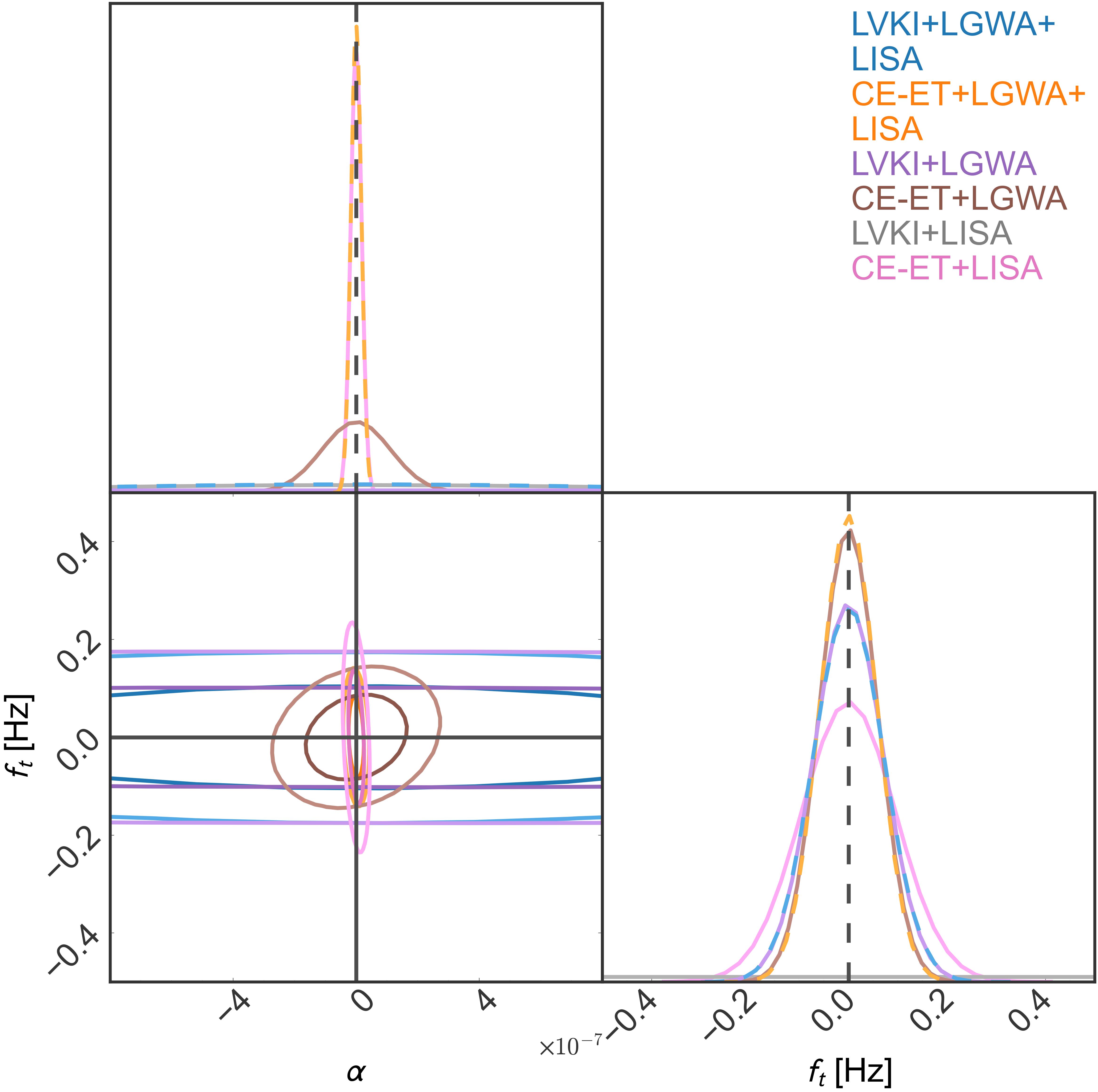}
		\caption{Posterior with $\alpha$ and $f_t$}
		\label{fig:GR1}
	\end{subfigure}
	\hfill
	\begin{subfigure}[b]{0.5\textwidth}
		\centering
		\includegraphics[width=9cm]{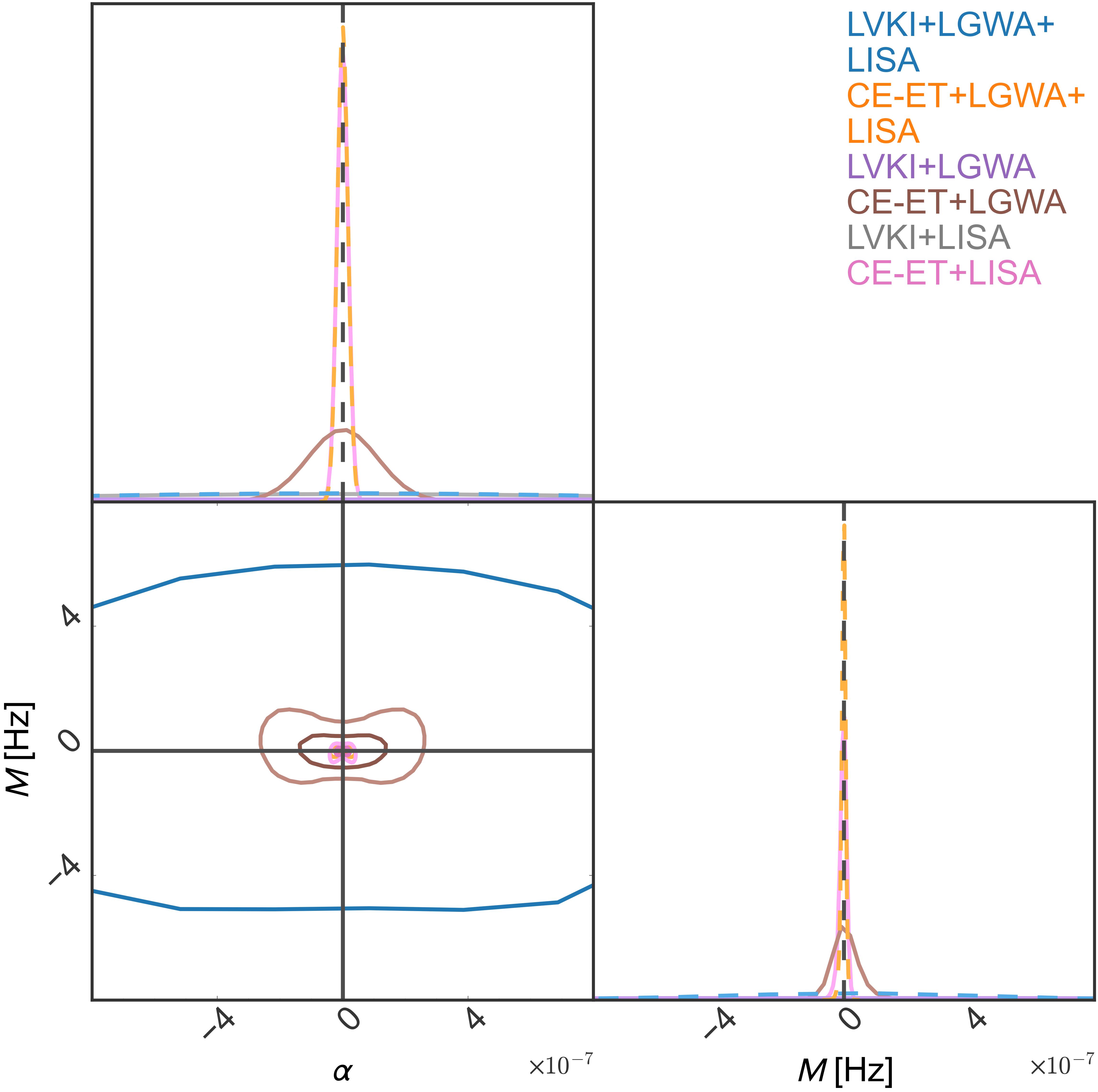}
		\caption{Posterior with $\alpha$ and $M$}
		\label{fig:GR2}
	\end{subfigure}
	
	\caption{Estimates for non-GR parameters for the fiducial case ($\alpha=0$, $f_t=0$, $M=0$) with various detector combinations. The estimation of $M$ is done as per equation \eqref{eq:M}. The fiducial values are $\alpha = 10^{-10}$, $f_t = 10^{-7}$ \si{\hertz} and  {$M = 6.235 \times 10^{-17}$ \si{\hertz}}. }
	\label{fig:GR}
\end{figure}

\begin{figure}
	\centering
	\includegraphics[width=15cm]{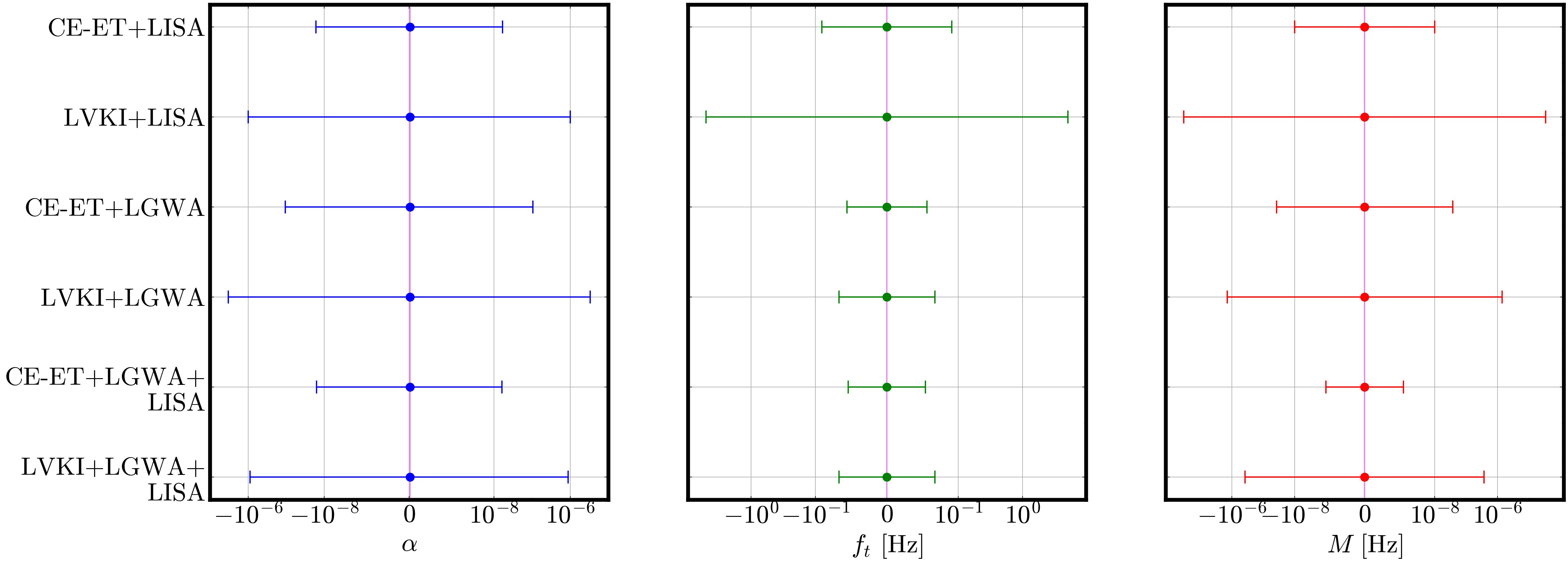}
	\caption{Error bars for various detector combinations of non-GR parameters in the fiducial ($\alpha=0$, $f_t=0$, $M=0$) scenario. The x-axis is made in symmetric log.}
	\label{fig:GR_errs}
\end{figure}

The results shown in figure \ref{fig:GR} focuses on the LVKI+LISA, CE-ET+LISA, LVKI+LGWA, CE-ET+LGWA, LVKI+LGWA+LISA and CE-ET+LGWA+LISA configuration . Error bars are shown in figure \ref{fig:GR_errs} and the values are mentioned in table \ref{tab:error bars}. From this it is clear that for the detector combinations considered, the zero transition frequency lies in the $1\sigma$ band. These results show that the multi-band observation by combining LISA+LGWA+CE-ET can provide the tightest constraints on the energy scale M, due to tight constraints on both $f_t$ and $\alpha$ (see the table \ref{tab:error bars}). 


Furthermore, from these results we find that CE-ET combinations perform better than LVKI. This is mainly because of better noise characteristics, especially in the low frequency regime close to $f_t$, which helps CE-ET see more effect of the perturbation and also helps in better estimation of $f_t$ with LISA than LVKI+LISA. The choice of IMBHs, whose $f_{ISCO}$ is of the order $10$ \si{\hertz}, which needs better low frequency noise characteristics, also plays a role. Furthermore we see that the addition of LGWA greatly improves the error bars, especially for LVKI and LISA combination. This is because of our choice of IMBHs as the mass range of black holes considered. These black holes have high sensitivity in deci-Hertz detectors, and hence LGWA greatly improves the bounds we get on the non-GR parameters for the null case we are considering. This demonstrates the key role of a deci-Hertz GW detector for precision measurement of signature of energy scale of EFT.

\section{Conclusions}
\label{Conclusions}
Understanding the source of cosmic acceleration is one of the key science questions in Physics. A primary avenue to answer this question relies on the accurate inference of the cosmological observables at the largest scale . GWs provide one such avenue to probe modified gravity theories which can explain cosmic acceleration by tiny distortion in the GW waveform due to frequency dependent dispersive effect . Such signature depends on the energy scale of the beyond GR theories and its measurement provides a direct way to discover such energy scales.

In this work, we demonstrate that with multi-band GW observations, especially those including a deci-Hertz detector like LGWA, we can put strong constraints on the dispersion of GWs, and on the energy scale of dark energy EFT and its coupling with gravity. The dispersion relation is modeled as per equation \eqref{eq8} with transition frequency $f_t = 0.26$ \si{\hertz}. We have chosen intermediate mass black holes for our analysis, due to their better sensitivity for various detectors and shorter timescale of evolution, making them ideal candidates for multi-band analysis. Our results show that with one year of observation, we will be able to detect dispersion $\frac{\delta c}{c}$ of the order $10^{-20}$ in a wide frequency range(from LISA band to LVKI/CE-ET band) with four detector combinations, namely CE-ET+LGWA and CE-ET+LGWA+LISA . From the constraint on the amount of dispersion and transition frequency, we are able to put the following relative constraints on the energy scale $M$ at $\frac{\delta c}{c} = 2 \times 10^{-20}$ , 

\begin{itemize}
	\item CE-ET+LGWA : $\frac{\sigma_{M}}{M} \approx 9.3\%$,
	\item CE-ET+LGWA+LISA : $\frac{\sigma_{M}}{M} \approx 8.6\%$.
\end{itemize}

Though we do not rule out the GR scenario for the fiducial values considered, LVKI+LGWA+LISA also has a stringent error bar of $\sigma_{M} = 4.7 \times 10^{-10}$ \si{\hertz}. With such constraints on the scale of EFT of dark energy and its coupling with gravity, we can put a strong bond on the parameters of alternate GR theories and even rule out some of the . Furthermore, by putting constraints on the speed of GWs over a wide frequency range, we are able to test GR at multiple length scale. At the present, we have excellent tests of GR at small length scales/high energy scale. However there could be deviations from GR at large length scales which can be probed by multi-band GW analysis discussed in this paper.

For getting good estimates, choosing a suitable source for doing multi-band analysis is key as the source should have good SNR in multiple detectors operating in various frequency bands, and also the timescale of evolution of the source through the various detector bands should be within the lifetime of the detector . Intermediate mass black holes, with their good SNR in multiple detectors(as shown in figure \ref{fig_sens}), and timescale of evolution in the order of a year going from LISA frequency band to LVKI/CE-ET frequency band, makes them the ideal source for doing multi-band analysis. With such sources, having good sensitivity in the frequency region $10^{-1}$ \si{\hertz} to $10^{1}$ \si{\hertz} is key as these sources merge at frequencies in this range. Hence having detectors deci-Hertz detectors such as LGWA, and even detectors such as CE-ET, which has good sensitivity in this frequency band is key for doing multi-band analysis. In our case, the fiducial value of transition frequency is $f_t = 0.26$ \si{\hertz}(in LGWA band), making the presence of LGWA even more important. But even if the transition frequency is not in LGWA band, as discussed in section \ref{Non-GR Case}, the presence of both LGWA and CE-ET greatly improves the error bars we obtain. Hence the presence of detector sensitive in the deci-Hertz \citep{Kawamura:2006up, Harms2021} region is crucial for doing multi-band analysis like the one presented in this paper.


%


%

In summary, the analysis presented in this paper highlights the sensitivity of the future detectors to dispersion in GWs, and shows that with a few sources having good SNR in multiple detectors, the dispersion in GWs can be probed to a very high accuracy over a wide range of frequency. With IMBHs being the best candidates for performing multi-band analysis, it is crucial to have a detector that can observe such sources with high sensitivity in the deci-Hertz frequency band, such as LGWA. In future, these joint observations can either discover or rule out dispersion of GW signal and can provide deep insight into the alternate theories of gravity.

\acknowledgments
This work is part of the $\langle \texttt{data|theory}\rangle$ \texttt{Universe-Lab} which is supported by the TIFR and the Department of Atomic Energy, Government of India.  AP and PD would like to thank NIUS program of HBCSE-TIFR funded by the Department of Atomic Energy, Govt. of India (Project No. RTI4001 . The authors are grateful for computational resources the 〈data|theory〉 Universe-Lab, supported by TIFR and the Department of Atomic Energy, Government of India. The authors also thank the LIGO-Virgo-KAGRA Scientific Collaboration for providing noise curve . We acknowledge the following Python packages in this work: NumPy\citep{numpy}, SciPy\citep{scipy}, Matplotlib\citep{matplotlib}, pyGTC\citep{pygtc}, SciencePlots\citep{SciencePlots}, BBHx\citep{michaelkatz2021,PhysRevD.102.023033,PhysRevD.105.044055}, GWFish\citep{Dupletsa2023} and Bilby\citep{Ashton2019}.

\bibliography{testing_GR}

\appendix
\section{Effective field theory calculations}
\label{Effective field theory calculations}

Here we solve the Lagrangian given in equation \eqref{eq7}  {in the limit of $\delta \phi \ll \langle\phi\rangle$ and $|\partial^2\phi | \ll M^2$}, that is the changes in the field are small and in low frequency limit. Since far away from source is where the non-GR effects add up, we can consider the  {spacetime as approximately flat (linearized gravity)}. Here we follow $\eta_{\mu\nu} = \mathrm{diag}(-1,+1,+1,+1)$ notation. In this limit, the Lagrangian becomes,
\begin{equation}
	\mathcal{L} = -\frac{1}{2}(\partial \phi)^2 + \frac{(\partial \phi)^4}{2 \Lambda^4} + \mathcal{O}\left(\frac{(\partial^2 \phi)^2}{M^2}\right),
\end{equation}
where $(\partial \phi)^2 = \partial_{\mu}\phi\partial^{\mu}\phi$.
The Euler-Lagrange equations are,
\begin{equation}
	\partial_{\mu}\left( \frac{\partial \mathcal{L}}{\partial(\partial_{\mu}\phi)} \right) - \frac{\partial \mathcal{L}}{\partial \phi} = 0.
\end{equation}
Here we have
\begin{equation}
	\frac{\partial \mathcal{L}}{\partial \phi} = 0
\end{equation}
and,
\begin{equation}
	\frac{\partial \mathcal{L}}{\partial(\partial_{\mu} \phi)} = \partial^{\mu}\phi \left( -1 + 2\frac{(\partial \phi)^2}{\Lambda^4} \right).
\end{equation}
After substitution to the Euler-Lagrange equations and further simplifications, we get,
\begin{equation}
	-\partial_{\mu}\partial^{\mu}\phi + 2\frac{(\partial_{\mu}\partial^{\mu}\phi)(\partial_{\nu}\phi \partial^{\nu}\phi)}{\Lambda^4} + 2\frac{\partial^{\mu} \phi \partial_{\mu}(\partial_{\nu}\phi\partial^{\nu}\phi)}{\Lambda^4} = 0.
\end{equation}
We then put in the limits of small perturbation to get,
\begin{align}
	\partial_{\mu}\phi\partial^{\mu}\phi &= -(\partial_t \phi^2) + (\nabla \phi)^2\\
	&= -\alpha^2\Lambda^4 - 2\alpha\Lambda^2\partial_{0}(\delta \phi) + \mathcal{O}((\delta \phi)^2),
\end{align}
and \begin{align}
	\partial^{\mu}\phi\partial_{\mu}(\partial_{\nu}\phi\partial^{\nu}\phi) = 2\alpha^2\Lambda^4\partial_{t}^2(\delta\phi)+\mathcal{O}((\delta\phi)^2).
\end{align}
Also, $\partial_{\mu}\partial^{\mu}\phi = \partial_{\mu}\partial^{\mu}\delta\phi$ since $\langle\phi\rangle$ is linear in time and independent of space coordinates. With these we get the following wave equation for the perturbations $\delta\phi$,
\begin{equation}
	\frac{\partial^2(\delta\phi)}{\partial^2 t} = \frac{1+2\alpha^2}{1+6\alpha^2} \nabla^2(\delta\phi).
\end{equation}
From the above equation the speed $c_{gw}$ can be inferred as,
\begin{align}
	c_{gw}^2 &= \frac{1+2\alpha^2}{1+6\alpha^2},\\
	&= 1 - \frac{4\alpha^2}{1+6\alpha^2}.
\end{align}
 {In the low $\alpha$ limit ($\alpha \ll 1$), we have 
\begin{equation}
    c_{gw}^2 = 1 - \frac{4\alpha^2}{1+2\alpha^2} \approx 1 - 4\alpha^2,
\end{equation}
and hence, 
\begin{equation}
        c_{gw} = \sqrt{1-4\alpha^2} \approx 1 - 2\alpha^2.
\end{equation}
}
\\
 {
Thus, deviation in the speed in low alpha limit is thus given as $\delta_g = 1-c_{gw} \approx 2\alpha^2$. Note that we use a scalar effective field theory to describe the dispersion in GW}.

\section{Mass model}
\label{Mass model}
A smoothened power law \citep{Karathanasis2023} is used for the mass distribution which is an extension of mass distribution for stellar mass black holes from LIGO and information from hierarchical merger . We do not include the Gaussian bump as there is no physical reason to include it.

The broken power law is defined as :
\begin{equation}
	P_l(x\mid\{\alpha,x_{min},x_{max}\}) \propto 
	\begin{cases}
		x^{-\alpha} & \text{if } x_{min}\leq x \leq x_{max}\\
		0 & \text{else}
	\end{cases}
	.
	\label{eq11}
\end{equation}
\\\\
The smoothing function is defined as :
\begin{equation}
	S(x\mid\{x_{min},\delta\}) = 
	\begin{cases}
		f(x-x_{min},\delta) & \text{if }x_{min}\leq x \leq x_{min}+\delta \\
		0 & \text{else}
	\end{cases}
	,
	\label{eq12}
\end{equation}
where
\begin{equation}
	f(x\mid\delta) = \left( 1+\exp{\left(\frac{\delta}{x}+\frac{\delta}{x-\delta}\right)} \right)^{-1}.
	\label{eq13}
\end{equation}
\\\\
From these the probability density function for the primary mass($m_1$) and mass ratio($q$) respectively are :
\begin{equation}
	P_{m_1}(m_1\mid\Lambda_{m_1}) =
	\begin{cases}
		P_l(m_1\mid\{\alpha,M_{min},M_{max}\})\hspace{1mm} S(m_1\mid\{M_{min},\delta_{m_1}\}) & \text{if  }M_{min}\leq m_1 \leq M_{min}+\delta_m\\
		P_l(m_1\mid\{\alpha,M_{min},M_{max}\}) & \text{if  }M_{min}+\delta_m < m_1 < M_{max} \\
		0 & \text{else}
	\end{cases}
	,
	\label{eq14}
\end{equation}
\vspace{0.6mm}
\begin{equation}
	P_{q}(q\mid\Lambda_{q}) =
	\begin{cases}
		P_l(q\mid\{\beta,q_{min},q_{max}\})\hspace{1mm}S(q\mid\{q_{min},\delta_{q}\}) & \text{if  }q_{min}\leq q \leq q_{min}+\delta_q\\
		P_l(q\mid\{\beta,q_{min},q_{max}\}) & \text{if  }q_{min}+\delta_q < q < q_{max} \\
		0 & \text{else}
	\end{cases}
	,
	\label{eq15}
\end{equation}
\vspace{1mm}
where $\Lambda_{m_1} = \{\alpha,M_{min},M_{max},\delta_m\}$ and $\Lambda_{q} = \{\beta,q_{min},q_{max},\delta_q\}$. 
\\\\
From this, we get the probability density distribution for the secondary mass as follows
\begin{equation}
	P_{m_2}(m_2 \mid m_1,\Lambda_{q}) = 
	\begin{cases}
		P_q(m_2/m_1,\Lambda_q) & \text{if } m_2/m_1 \leq 1 \\
		0 & \text{otherwise}
	\end{cases}
	.
	\label{eq16}
\end{equation}
\\
Table \ref{Tab1} gives the fiducial values of these parameters.
\begin{table}[H]
	\caption{Fiducial value of parameters in mass model}
	\begin{tabular}{|p{3cm}|p{3cm}||p{3cm}|p{3cm}|}
		\hline
		\multicolumn{2}{|c||}{Parameters for primary mass distribution} &\multicolumn{2}{c|}{Parameters for mass ratio distribution} \\
		\hline
		$\alpha$ & $3.4$ & $\beta$ & $0.8$ \\
		$M_{min}$ & $70$ & $q_{min}$ & $0$ \\
		$M_{max}$ & $500$ & $q_{max}$ & $1$ \\
		$\delta_{m_1}$ & $200$ & $\delta_q$ & $0.4$\\
		\hline
	\end{tabular}
	\label{Tab1}
\end{table}


\section{Redshift model}
\label{Redshift model}

The black hole population in this analysis is taken as IMBH population present in the disks of AGN . Thus black hole merger rate and population density are proportional to the AGN density $n_{AGN}(z)$ (where $z$ is redshift . The AGN density, and hence the IMBH merger rate and density, can be found from the AGN luminosity function $\phi_L(L,z)$ (where $L$ is luminosity), which is given as \citep{Yang2020}
\begin{equation}
	\phi_L(L,z) = \frac{\phi_\star(z)}{\left(\frac{L}{L_\star(z)}\right)^{\gamma_1(z)}+\left(\frac{L}{L_\star(z)}\right)^{\gamma_2(z)}},
\end{equation}
where
\begin{equation*}
	\gamma_1(z) = a_0 T_0(1+z)+a_1 T_1(1+z) + a_2 T_2(1+z) ,
\end{equation*}
\begin{equation*}
	\gamma_2(z) = \frac{2 b_0}{\left(\frac{1+z}{3}\right)^{b_1}+\left(\frac{1+z}{3}\right)^{b_2}} ,
\end{equation*}
\begin{equation*}
	\log(L_\star(z)) = \frac{2 c_0}{\left(\frac{1+z}{3}\right)^{c_1}+\left(\frac{1+z}{3}\right)^{c_2}} ,
\end{equation*}
\begin{equation*}
	\log(\phi_\star(z)) = a_0 T_0(1+z) + a_1 T_1(1+z),
\end{equation*}
where $T_n$ in the $n^{th}$ Chebyshev polynomial. The values for the 11 fit parameters are 
$\{a_1,a_2,a_3;b_1,b_2,b_3; \\c_1,c_2,c_3;d_1,d_2\} = \{0.8396,-0.2519,
0.0198;2.5432,-1.0528,1.1284;13.0124,-0.5777,0.4545;\\
-3.5148,-0.4045\}$.

The density of AGN is given by
\begin{equation}
	n_{AGN}(z) = \int_{L_{min}}^{L_{max}}\phi_L(L,z)\text{ } \mathbf{d}L ,
\end{equation}
where $L_{min} = 10^{41}\text{ erg/s}$ and $L_{max} = 3.15 \times 10^{14} L_{\odot}$ . Here $L_{\odot}$ is solar luminosity.
\\
The mass of the supermassive black holes $M_{\bullet}$ is related to the accretion rate given by
\begin{equation}
	\frac{M_\bullet}{M_\odot} = 3.17 \times 10^{-17}\frac{1-\epsilon}{\dot{m}} ,
\end{equation}
where $\epsilon$ is the radiation efficiency and $\dot{m}$ is the accretion rate which is given by 
\begin{equation}
	\dot{m} = (1-\epsilon) \lambda
\end{equation}
where \( \lambda = L/L_{edd}\) is the Eddington ratio.

The Eddington ratio distribution is given by 
\begin{equation}
	P(\lambda \mid L,\epsilon) = f_{uno} P_1(\lambda \mid z) + f_{obs} P_2(\lambda \mid z) ,
\end{equation}
where $f_{uno}$ and $f_{obs}$ are the fraction of unobscured(Type-1) and obscured(Type-2) AGN. $P_1$ and $P_2$ are the Eddington ratio distributions for type-1 and type-2 distributions respectively.
\\\\
The average black hole merger rate is given by 
\begin{equation}
	\Gamma(\dot{m}) = \frac{N_{disk}(\dot{m}) -1 + \exp(-N_{disk}(\dot{m}))}{\tau_{AGN}} ,
\end{equation}
where $\tau_{AGN} = 10^7 \text{ yr}$ is the AGN lifetime and $N_{disk}$ is the mean number of stellar black holes in the AGN disk, which has been assumed as a univariate function of $\dot{m}$ and obtained a power-law fit given as 
\begin{equation}
	N_{disk}(\dot{m}) = 5.5 \dot{m}^{1/3} .
\end{equation}

Combining all the above results, we obtain the IMBH merger rate as
\begin{equation}
	R_{AGN}(z) = \int_{L_{min}}^{L_{max}} \mathbf{d}\log{L}\text{  } \phi_L(L,z)\int_{\lambda_1}^{1} \mathbf{d}\lambda\text{  } \Gamma(\dot{m}) P(\lambda \mid L,z) ,
\end{equation}
where $L_{min} = 10^{41}\text{ erg/s}$,  $L_{max} = 3.15 \times 10^{14} L_{\odot}$ and $\lambda_1$ lie in range $10^{-4}$ to $10^{-2}$.
\end{document}